\documentclass[aps,twocolumn,superscriptaddress]{revtex4-1}
\usepackage{soul}
\usepackage[usenames,dvipsnames]{color}
\usepackage{amsmath}
\usepackage{amssymb}
\usepackage{graphicx}
\usepackage{tabularx}
\usepackage{color}
\usepackage[dvipdfm,bookmarks=true,colorlinks=true,linkcolor=blue,urlcolor=blue,
citecolor=blue]{hyperref}

\begin{document}

\title{Finite-Temperature Charge Dynamics and the Melting of the Mott Insulator}

\author{Xing-Jie Han}
\thanks{These authors contributed equally to this study.}
\affiliation{Beijing National Laboratory for Condensed Matter Physics and
\\ Institute of Physics, Chinese Academy of Sciences, Beijing 100190,
China}
\affiliation{Institut f\"ur Theoretische Festk\"orperphysik, RWTH Aachen
University, 52056 Aachen, Germany}
\author{Chuang Chen}
\thanks{These authors contributed equally to this study.}
\affiliation{Beijing National Laboratory for Condensed Matter Physics and
\\ Institute of Physics, Chinese Academy of Sciences, Beijing 100190,
China}
\affiliation{University of the Chinese Academy of Sciences, Beijing 100049,
China}
\author{Jing Chen}
\affiliation{Beijing National Laboratory for Condensed Matter Physics and
\\ Institute of Physics, Chinese Academy of Sciences, Beijing 100190,
China}
\author{Hai-Dong Xie}
\affiliation{Beijing National Laboratory for Condensed Matter Physics and
\\ Institute of Physics, Chinese Academy of Sciences, Beijing 100190,
China}
\author{Rui-Zhen Huang}
\affiliation{Beijing National Laboratory for Condensed Matter Physics and
\\ Institute of Physics, Chinese Academy of Sciences, Beijing 100190,
China}
\author{Hai-Jun Liao}
\affiliation{Beijing National Laboratory for Condensed Matter Physics and
\\ Institute of Physics, Chinese Academy of Sciences, Beijing 100190,
China}
\author{B. Normand}
\affiliation{Neutrons and Muons Research Division, Paul Scherrer Institute,
CH-5232 Villigen PSI, Switzerland}
\author{Zi Yang Meng}
\affiliation{Beijing National Laboratory for Condensed Matter Physics and
\\ Institute of Physics, Chinese Academy of Sciences, Beijing 100190,
China}
\affiliation{CAS Center of Excellence in Topological Quantum Computation
and School of Physical Sciences, University of the Chinese Academy of Sciences,
Beijing 100190, China}
\author{Tao Xiang}
\affiliation{Beijing National Laboratory for Condensed Matter Physics and
\\ Institute of Physics, Chinese Academy of Sciences, Beijing 100190,
China}
\affiliation{Collaborative Innovation Center of Quantum Matter, Beijing
100190, China}

\begin{abstract}
The Mott insulator is the quintessential strongly correlated electronic
state. We obtain complete insight into the physics of the two-dimensional
Mott insulator by extending the slave-fermion (holon-doublon) description
to finite temperatures. We first benchmark its predictions against
state-of-the-art quantum Monte Carlo simulations, demonstrating quantitative
agreement. Qualitatively, the short-ranged spin fluctuations both induce
holon-doublon bound states and renormalize the charge sector to form the
Hubbard bands. The Mott gap is understood as the charge gap renormalized
downwards by these spin fluctuations. As temperature increases, the Mott
gap closes before the charge gap, causing a pseudogap regime to appear
naturally during the melting of the Mott insulator.
\end{abstract}

\maketitle

\section{Introduction}
\label{sintro}

The Mott insulator \cite{Mott-1937} and its associated metal-insulator
transition (MIT) \cite{Mott-1949,Mott-1956,Imada-1998} are phenomena
generic to strongly correlated electron systems. The discovery
\cite{Bednorz-1986} of high-$T_{c}$ superconductivity in a class of
quasi-two-dimensional (quasi-2D) doped Mott insulators \cite{Lee-2006}
triggered an enduring experimental and theoretical quest to understand the
many anomalous properties of cuprates, including the strange metal, the
pseudogap \cite{Damascelli-2003,Norman-2005,Sawatzky-2008}, and indeed the
superconductivity itself, in a complete and correct description of the Mott
insulator.

In Mott's original proposal \cite{Mott-1990}, the insulating state arises
due to the strong on-site Coulomb interaction, $U$, and has no explicit
relation to symmetry-breaking (usually magnetic order). Hubbard
\cite{Hubbard-1963} obtained the incoherent upper and lower Hubbard bands
and considered the interaction-driven MIT, while Brinkman and Rice associated
the MIT with a diverging quasiparticle mass \cite{Brinkman-1970}. These seminal
results do not, however, include the spin fluctuations and their influence
on the charge dynamics. In experiment, most Mott insulators possess
antiferromagnetic (AFM) long-range order at low temperatures \cite{Imada-1998}.
In 2D, where order is forbidden at $T > 0$, the low-energy physics is dominated
by short-ranged spin fluctuations \cite{Huscroft-2001,Moukouri-2001,Kyung-2006,
Park-2008,Sordi-2010,Gunnarsson-2015}. On the scale of $U$, charge fluctuations
create empty sites (holons) and doubly-occupied sites (doublons), whose
tendency to form bound states has been proposed as the key to the high-energy
physics of the Mott insulator \cite{Castellani-1979,Kaplan-1982,Capello-2005,
Yokoyama-2006,Phillips-2010,Sen-2014,Peter-2015,Han-2016}. Clearly a full
description requires a proper account of both charge and spin fluctuations
\cite{Kotliar-1986}. While progress has been made in this direction through
the development of many sophisticated numerical methods \cite{LeBlanc-2015},
a physical understanding remains far from complete.

Here we provide this insight by treating the half-filled Hubbard model within
a slave-fermion formulation, in which the charge degrees of freedom are
represented by fermionic holons and doublons, while the spin degrees of
freedom are bosonic. To demonstrate that this framework provides quantitatively
accurate resuls at the mean-field level for all temperatures, we perform
detailed determinantal quantum Monte Carlo (QMC) simulations for benchmarking
purposes. The key capability of our analytical approach is that it treats both
the low- (spin) and high-energy (charge) degrees of freedom consistently,
thereby capturing qualitative effects that arise due exclusively to their
interplay, which to date have been accessible only by numerical methods.

By inspecting the physical content of the holon-doublon description, we are
able to unveil the phenomenology of the Mott-insulating state. Our primary
conclusions are as follows. Long-ranged AFM order is not required, because
short-ranged spin fluctuations induce holon-doublon bound states. These
fluctuations renormalize the charge sector to produce a Mott gap that is
smaller than the charge gap. This reconstruction of the electronic states
produces a quasiparticle, the ``generalized spin polaron,'' at the
Hubbard-band edges, while most of the composite states lie higher in energy.
The differing thermal evolution of the two gaps explains the origin of the
pseudogap in the Mott insulator, which is a generic property at half-filling
as well as at finite doping. The mixing of energy scales long observed to be
at the heart of the complications endemic to a theoretical treatment of the
Mott-insulating state is shown to be a ``leverage'' effect intrinsic to the
convolution of charge and spin sectors that forms the reconstruction process.

The structure of this article is as follows. In Sec.~\ref{shmsf} we
introduce the slave-fermion description we use to analyze the Hubbard model
at finite temperatures. In Sec.~\ref{sec:QMC} we present details of our QMC
techniques and apply these to benchmark our holon-doublon calculations. Having
achieved a quantitative validation of its relevance, in Sec.~\ref{sccs} we use
the slave-fermion framework to deconstruct the single-particle excitations of
the Mott insulator into their charge and spin components and to investigate
their reconstruction over the full range of temperatures. In Sec.~\ref{si} we
discuss the emergence of the pseudogap regime from the slave-fermion analysis
and extract the intrinsic physics of the Mott insulator in terms of
quasiparticle reconstruction and energy-scale leverage. Section \ref{scr}
contains a short summary and perspective.

\section{Hubbard model: Slave-fermion formalism when $T > 0$}
\label{shmsf}

The Hamiltonian for the one-band Hubbard model is
\begin{equation}
H = - t \sum_{\langle i,j \rangle \sigma} c_{i\sigma}^{\dagger} c_{j\sigma} + U \sum_{i}
(n_{i\uparrow} - {\textstyle \frac12})(n_{i\downarrow } - {\textstyle \frac12}),
\label{ehm}
\end{equation}
where $c_{i\sigma }^{\dagger }$ creates an electron with spin $\sigma$ on site $i$
and $\langle i,j \rangle$ indicates only nearest-neighbor hopping, whose
amplitude, $t = 1$, sets the unit of energy. In the large-$U$ limit, the
half-filled Hubbard model can be mapped to the AFM Heisenberg model
\cite{Auerbach-1994}, $H_{S} = J \sum_{\langle i,j \rangle} \mathbf{S}_{i} \cdot
\mathbf{S}_{j}$, and we assume that the spin dynamics are governed by this
model also at finite $U$, taking $J = 4t^{2}/U$.

The application of slave-particle decompositions to the Hubbard model
[Eq.~(\ref{ehm})] has a long history, which is reviewed briefly in
Ref.~\cite{Han-2016}. While in general the attribution of fermionic or
bosonic statistics to the charge and spin components of the electron is
arbitrary, for studies constrained to and about the mean-field level it was
found shortly after the discovery of high-temperature superconductivity in
cuprates \cite{Bednorz-1986} that the slave-fermion decomposition is more
appropriate for the low-doped regime dominated by magnetic order (or
correlations), while the superconducting and strange-metal regime is better
described by the slave-boson decomposition \cite{Lee-2006}. For the present
purposes, the magnetic fluctuations of the $S = 1/2$ square-lattice
antiferromagnet are significantly better described by Schwinger bosons
\cite{Arovas-1988} and the charge sector, which is expected to undergo
a holon-doublon binding analogous to the electron binding in
Bardeen-Cooper-Schrieffer (BCS) superconductivity, by fermionic statistics.

Thus we employ a slave-fermion formalism \cite{Yoshioka-1989} in which the
electron operator is expressed as
\begin{equation}
c_{i\sigma } = s_{i\overline{\sigma }}^{\dagger} d_{i} + \sigma e_{i}^{\dagger} s_{i\sigma },
\label{esfd}
\end{equation}
where $e_{i}$ and $d_{i}$ are fermionic operators denoting the charge degrees
of freedom, respectively holons and doublons, and $s_{i\sigma}$ are bosonic
operators describing the spins, with $\sigma = 1 (-1)$ for spin $\uparrow$\
($\downarrow$). The physical Hilbert space is established by the constraint
$d_{i}^{\dagger } d_{i} + e_{i}^{\dagger} e_{i} + \sum_{\sigma} s_{i\sigma }^{\dagger}
s_{i\sigma } = 1$, which for an analytical treatment is satisfied only globally
rather than locally.

The Hubbard model (\ref{ehm}) now takes the form
\begin{align}
H & = - t \sum_{i,\delta ,\sigma} [(d_{i+\delta}^{\dagger} d_{i} - e_{i+\delta}^{\dagger}
e_{i}) s_{i,\sigma}^{\dagger} s_{i+\delta,\sigma} + \text{h.c.}] \notag \\
& \quad - t \sum_{i,\delta,\sigma} [(d_{i}^{\dagger} e_{i+\delta}^{\dagger} + e_{i}^{\dagger}
d_{i+\delta}^{\dagger}) \sigma s_{i,\bar{\sigma}} s_{i+\delta,\sigma} + \text{h.c.}] \notag
\\ & \quad + {\textstyle \frac{1}{2}} U \sum_{i} (d_{i}^{\dagger} d_{i}
 + e_{i}^{\dagger} e_{i} - {\textstyle \frac{1}{2}}),  \label{HTU}
\end{align}
where $\delta $ denotes lattice vectors $(a,0)$ and $(0,a)$, with $a$ the
lattice constant. The first two lines make clear that the spin and charge
degrees of freedom are intertwined, whence AFM fluctuations cause a
holon-doublon pairing interaction. In our previous work \cite{Han-2016},
we studied the physical content of this formalism at $T = 0$, where the
long-ranged magnetic order is described by the condensation of a single
bosonic spin operator, $\langle s_{i,\sigma}^{\dagger} \rangle \neq 0$. At
$T > 0$, only short-range AFM fluctuations are present and these are well
described at the mean-field level by two-operator condensation of the form
$\sum_{\sigma} \langle \sigma s_{i,\bar{\sigma}} s_{i+\delta,\sigma} \rangle \neq 0$
(while $\langle s_{i,\sigma }^{\dagger } \rangle = 0$) on the bonds connecting
all sites $i$ to their nearest neighbors. In a recent study of the doped
Hubbard model \cite{rwscsgf,rscwfgs}, which also used an ansatz with
fermionic charge, this short-range-correlated spin state was described
by an SU(2) gauge theory with a finite Higgs (amplitude) field but no
orientational order.

Taking the bosonic spin degrees of freedom, $s_{i}$, to be governed
by the Heisenberg model, we follow the treatment of Arovas and Auerbach
\cite{Arovas-1988}. Replacing $\sum_{\sigma} \langle \sigma s_{i,\bar{\sigma}}
s_{i+\delta,\sigma} \rangle$ by its mean value decouples the second line of
Eq.~(\ref{HTU}), allowing us to calculate the holon and doublon Green
functions within the self-consistent Born approximation (SCBA) \cite{Han-2016}.
By introducing the bond operator
\begin{equation}
Q_{i,\delta} = s_{i,\uparrow} s_{i+\delta,\downarrow} - s_{i,\downarrow} s_{i+\delta,\uparrow },
\end{equation}
one may reformulate the Heisenberg model as
\begin{equation}
H_{S} = - {\textstyle \frac{1}{2}} J \sum_{i,\delta} (Q_{i,\delta }^{\dag }Q_{i,\delta }
 - {\textstyle \frac{1}{2}}).
\label{eq:Hs}
\end{equation}
We take the mean-field parameter to be uniform and static,
\begin{equation}
Q = - {\textstyle \frac{1}{2}} J \langle s_{i,\uparrow} s_{i+\delta,\downarrow}
 - s_{i,\downarrow} s_{i+\delta,\uparrow} \rangle
\end{equation}
for all $i$ and $\delta$, and release the constraint on the slave-boson
sector, $s_{i\uparrow}^{\dag} s_{i\uparrow} + s_{i\downarrow}^{\dag} s_{i\downarrow} = 1$
\cite{Arovas-1988}, replacing it by the constraint $d_{i}^{\dag} d_{i} +
e_{i}^{\dag} e_{i} + \sum_{\sigma} s_{i\sigma}^{\dag} s_{i\sigma} = 1$ appropriate to
the full slave-fermion problem \cite{Han-2016}. The constraint acts to
provide an additional and self-consistent coupling of the spin and charge
degrees of freedom. In principle, the corresponding two-operator expectation
value $P = \langle s_{i,\uparrow}^\dag s_{i+\delta,\uparrow} + s_{i,\downarrow}^\dag
s_{i+\delta,\downarrow} \rangle$ is also finite in the coupled problem, but we find
from the three-parameter mean-field solution that its value is sufficiently
small, at all temperatures, for its neglect to be fully justified in the
treatment to follow.

The mean-field Hamiltonian can be expressed as
\begin{eqnarray}
H_{S} & = & \sum_{\mathbf{k}} ( \! \begin{array}{cc}
s_{\mathbf{k},\uparrow}^{\dag} & s_{-\mathbf{k},\downarrow} \end{array} \! )
\! \left( \!\! \begin{array}{cc} \lambda & z Q  \eta_{\mathbf{k}} \\
z Q \eta_{\mathbf{k}}^{\ast} & \lambda \end{array} \!\! \right) \!\!
\left( \!\! \begin{array}{c} s_{\mathbf{k},\uparrow} \\ s_{-\mathbf{k},\downarrow}^{\dag }
\end{array} \!\! \right) \nonumber \\ & & + \frac{N z |Q|^{2}}{J} - 2 \lambda
N + \lambda \sum_{i} (d_{i}^{\dag} d_{i} + e_{i}^{\dag} e_{i}),
\end{eqnarray}
where $z = 4$ is the coordination number and $\eta_{\mathbf{k}} = {\textstyle
\frac{1}{2}} i (\sin k_{x} + \sin k_{y})$.
The Bogoliubov transformation
\begin{equation*}
\left( \!\! \begin{array}{c} s_{\mathbf{k},\uparrow} \\ s_{-\mathbf{k},\downarrow}^{\dag }
\end{array} \!\! \right) = \left( \!\! \begin{array}{cc} u_{\mathbf{k}} &
v_{\mathbf{k}} \\ v_{\mathbf{k}}^{\ast} & u_{\mathbf{k}}^{\ast} \end{array} \!\! \right)
\! \left( \!\! \begin{array}{c} \alpha_{\mathbf{k}} \\ \beta_{-\mathbf{k}}^{\dag}
\end{array} \!\! \right) \! ,
\end{equation*}
with
\begin{eqnarray}
|u_{\mathbf{k}}|^{2} & = & \frac{1}{2} + \frac{\lambda}{2 \Omega_{\mathbf{k}}}, \;\;
\;\; |v_{\mathbf{k}}|^{2} =  - \frac{1}{2} + \frac{\lambda}{2 \Omega_{\mathbf{k}}},
\label{euv}
\\ u_{\mathbf{k}} v_{\mathbf{k}} & = & - \frac{z Q \eta _{\mathbf{k}}}{2
\Omega_{\mathbf{k}}}, \;\;\;\; \Omega_{\mathbf{k}} = \sqrt{\lambda^{2} - 4Q^{2}
(\sin k_{x} + \sin k_{y})^{2}}, \nonumber
\end{eqnarray}
diagonalizes the Hamiltonian to yield the form
\begin{eqnarray}
H_{S} & = & \sum_{\mathbf{k}} \Omega_{\mathbf{k}} \alpha_{\mathbf{k}}^{\dag}
\alpha_{\mathbf{k}} + \sum_{\mathbf{k}} \Omega_{\mathbf{k}} \beta_{\mathbf{k}}^{\dag}
\beta_{\mathbf{k}} + \sum_{\mathbf{k}} \Omega_{\mathbf{k}} \nonumber \\ & &
 + \lambda \sum_{i} (d_{i}^{\dagger} d_{i} + e_{i}^{\dagger} e_{i}) + \frac{NzQ^{2}}{J}
 - 2 N \lambda,
\end{eqnarray}
where $\lambda $ is the Lagrange multiplier associated with the constraint.
The mean-field equations for any temperature, $T$, are given by
\begin{gather}
\frac{J}{N} \sum_{\mathbf{k}} \frac{z (\sin k_{x} + \sin k_{y})^{2}}{\Omega
_{\mathbf{k}}} \left( n_{\mathbf{k}} + \frac{1}{2} \right) = 1 \\ \frac{1}{N}
\! \sum_{\mathbf{k}} \! \frac{\lambda}{\Omega_{\mathbf{k}}} \! \left( \!
n_{\mathbf{k}} + \frac{1}{2} \right) = 1 - \frac{1}{2N} \! \sum_{i} \!
(d_{i}^{\dag} d_{i} + e_{i}^{\dag} e_{i}),
\end{gather}
where $n_{\mathbf{k}} = 1/(e^{\Omega_{\mathbf{k}}/T} - 1)$ is the Bose
distribution function. Self-consistent solution of these equations yields
temperature-dependent mean-field parameters, $\lambda(T)$ and $Q(T)$, whose
effect is to increase the excitation gap of the effective spin dispersion
relation of the thermally disordered magnetic system. It is important to
note that the gap in the spin spectrum remains significantly smaller than
$T$ at all relevant temperatures \cite{Arovas-1988}.

To combine the spin degrees of freedom with the charge, the mean-field
solution for the Heisenberg model is substituted into Eq.~(\ref{HTU}).
The most important term is the replacement of $(s_{i,\downarrow} s_{i+\delta,
\uparrow} - s_{i,\uparrow} s_{i+\delta,\downarrow })$ in the quadratic decoupling of
the second line by its mean value, $2Q/J$. Together with the third line,
this term forms an effective unperturbed Hamiltonian for the charge dynamics,
while the remaining terms describe interactions. With this separation,
Eq.~(\ref{HTU}) can be expressed as
\begin{equation}
H = \sum_{\mathbf{k}} \psi_{\mathbf{k}}^{\dag} \tilde{\varepsilon}_{\mathbf{k}}
\psi_{\mathbf{k}} + \sum_{\mathbf{k},\mathbf{q,l}} \psi_{\mathbf{k}}^{\dag} M(\mathbf{k},
\mathbf{q,l}) \psi_{\mathbf{k}-\mathbf{q+l}},
\label{SCBAHubbard}
\end{equation}
where $\psi_{\mathbf{k}}^{\dagger } = (d_{-\mathbf{k}}^{\dag},e_{\mathbf{k}})$ is the Nambu
spinor for the charge degrees of freedom,
\begin{equation}
\tilde{\varepsilon}_{\mathbf{k}} = \left( \!\! \begin{array}{cc}
U/2 & 2 t z Q \eta_{\mathbf{k}}/J \\ - 2 t z Q \eta_{\mathbf{k}}/J & - U/2
\end{array} \!\! \right) \! ,  \label{Dis}
\end{equation}
and
\begin{equation}
M(\mathbf{k},\mathbf{q,l}) = - \frac{tz}{N} \sum_{\sigma} \left( \!\!
\begin{array}{cc} \gamma_{\mathbf{k+l}} & 0 \\ 0 & \gamma_{\mathbf{k-q}} \end{array}
\!\! \right) s_{\mathbf{q},\sigma}^{\dag} s_{\mathbf{l},\sigma},
\end{equation}
in which $\gamma_{\mathbf{k}} = {\textstyle \frac{1}{2}} (\cos k_{x} + \cos k_{y})$.
The first term of Eq.~(\ref{SCBAHubbard}) describes the charge dynamics in the
absence of spin renormalization, with holon-doublon binding appearing in
the off-diagonal part of the matrix. The second term incorporates all the
interactions between the charge and spin degrees of freedom, which in contrast
to the $T = 0$ case \cite{Han-2016} contains two spin bosons and requires a
sum over three free momenta.

\begin{figure}[t]
\includegraphics[width=\columnwidth]{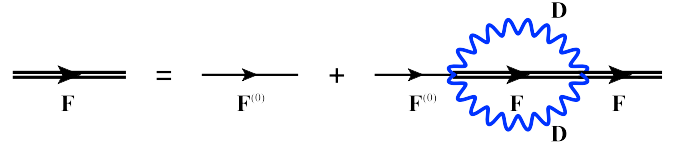}
\caption{Feynman diagrams for the self-consistent Born approximation.
Fermion (holon-doublon) and boson (magnon) propagators are represented
respectively by the straight and wavy lines.}
\label{fig:Feynman}
\end{figure}

We define the full charge, or holon-doublon, Matsubara Green function as
\begin{equation}
\mathbf{F} (\mathbf{k},\tau) = - \langle T_{\tau } \psi_{\mathbf{k}}(\tau) \psi
_{\mathbf{k}}^{\dag} (0) \rangle  \label{emgf}
\end{equation}
and calculate this within the SCBA. The corresponding Feynman diagrams, shown
in Fig.~\ref{fig:Feynman}, are the bare term, $\mathbf{F}^{(0)}$, and the first
loop, in which the magnon Green function is also a $2\times 2$ matrix,
\begin{equation*}
D(\mathbf{k},\tau) \! = \! - \! \left( \!\! \begin{array}{cc} \langle T_{\tau}
s_{\mathbf{k,\uparrow}}(\tau) s_{\mathbf{k,\uparrow}}^{\dag}(0) \rangle & \langle
T_{\tau} s_{-\mathbf{k,\downarrow }}^{\dag} (\tau) s_{\mathbf{k,\uparrow}}^{\dag} (0)
\rangle \! \\ \langle T_{\tau} s_{\mathbf{k,\uparrow}} (\tau) s_{-\mathbf{k,\downarrow}}
(0)\rangle & \langle T_{\tau} s_{-\mathbf{k,\downarrow}}^{\dag} (\tau)
s_{-\mathbf{k,\downarrow }} (0) \rangle \end{array} \!\! \right) \!.
\label{edgf}
\end{equation*}
At this level we obtain the self-consistent Dyson equation for the Matsubara
Green function of the charge sector,
\begin{equation}
\mathbf{F} (\mathbf{k},i\omega_{n}) = \frac{1}{i \omega_{n}
 - \tilde{\varepsilon}_{\mathbf{k}} - \mathbf{\Sigma} (\mathbf{k},i\omega_{n})},
\label{Charge}
\end{equation}
whence the retarded Green function is obtained by the analytic continuation
$i \omega_{n} \rightarrow \omega + i\eta $. This $\eta$ term denotes a
broadening of the peaks in the spectral response and is set to $\eta = 0.1$
throughout our calculations; a smaller value would be of little physical
meaning because of the finite system sizes in the calculations to follow.

We comment that, despite the simplicity of the AFM Heisenberg model,
there is no exact solution for the $S = 1/2$ case on the square lattice
\cite{Manousakis-1991}. The study of the two-dimensional (2D) quantum AFM
Heisenberg model is of great importance in its own right as a fundamental
problem in quantum magnetism. To date, the most definitive analytical results
for the low-temperature regime were obtained by two-loop renormalization-group
calculations on the quantum nonlinear $\sigma$ model (NL$\sigma$M)
\cite{Chakravarty-1989}. It has also been shown \cite{Schulz-1990,Dupuis-2004}
that spin fluctuations in the 2D Hubbard model at low temperature can be
described by the quantum NL$\sigma$M for any value of the Coulomb repulsion,
$U$. The accuracy of these results notwithstanding, an integration of the
NL$\sigma$M into the present analysis is not straightforward, and we will
show that the holon-doublon framework with mean-field decoupling is already
sufficient to gain semi-quantitative accuracy.

\section{Benchmarking SCBA: Quantum Monte Carlo}
\label{sec:QMC}

We investigate the half-filled 2D Hubbard model by determinantal QMC.
The quartic term in Eq.~(\ref{ehm}), $U (n_{i\uparrow} - {\textstyle
\frac{1}{2}})(n_{i\downarrow} - {\textstyle \frac{1}{2}})$, is decoupled by
Hubbard-Stratonovich transformation to a form quadratic in $(n_{i\uparrow}
 - n_{i\downarrow}) = (c^{\dagger}_{i\uparrow} c_{i\uparrow} - c^{\dagger}_{i\downarrow}
c_{i\downarrow})$ \cite{BSS-1981,Hirsch-1983,Hirsch-1985}, which introduces an
auxiliary Ising field on each lattice site. The QMC procedure obtains the
partition function of the underlying Hamiltonian in a path-integral
formulation in a space of dimension $N = L \times L$ and an imaginary time
$\tau$ up to $\beta = 1/T$. All of the physical observables are measured
from the ensemble average over the space-time ($N \beta$) configurational
weights of the auxiliary fields. As a consequence, the errors within the
process are well controlled; specifically, the $(\Delta \tau)^{2}$ systematic
error from the imaginary-time discretization, $\Delta \tau = \beta/M$, is
controlled by the extrapolation $M \to \infty$ and the statistical error
is controlled by the central-limit theorem (simply put, the larger the
number of QMC measurements, the smaller the statistical error).

The QMC algorithm is based on Ref.~\cite{BSS-1981} and has been refined by
including global moves \cite{Scalettar-1991} to improve ergodicity and delay
updating of the fermion Green function, which increases the efficiency of the
QMC sampling. Details concerning the QMC simulation code are provided in
Ref.~\cite{quest}. We have performed simulations for system sizes
$L = 4$, 8, 10, 12, 14, and 16. The interaction, $U$, is varied from $2$ to
$12$ in units of the hopping strength, which is set to $t = 1$, and for each
$U$ we simulate temperatures from $T = 0.0625$ to $1$ (inverse temperatures
$\beta = 1$ to $16$).

\begin{figure}[t]
\includegraphics[width=0.95\columnwidth]{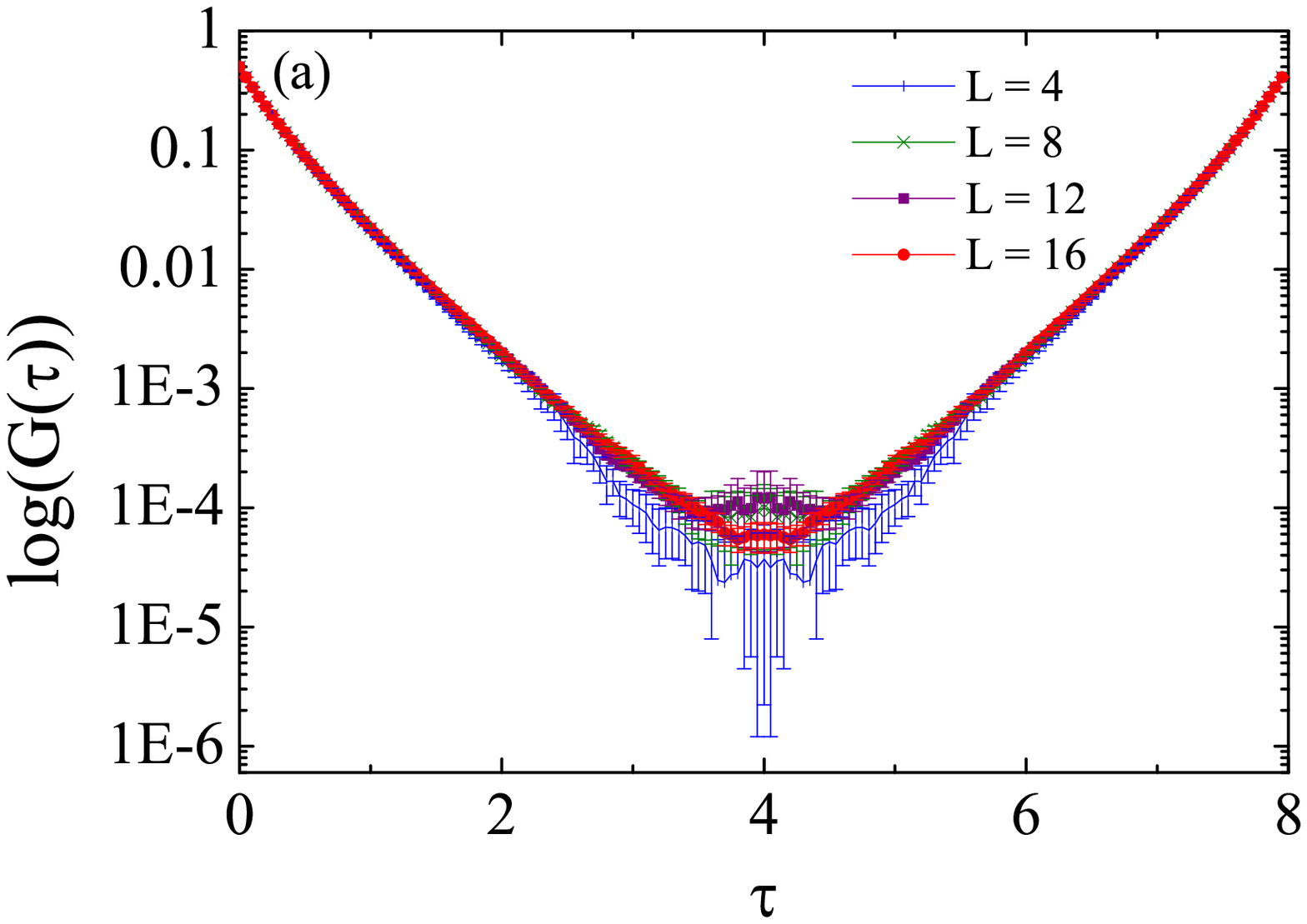}
\includegraphics[width=0.95\columnwidth]{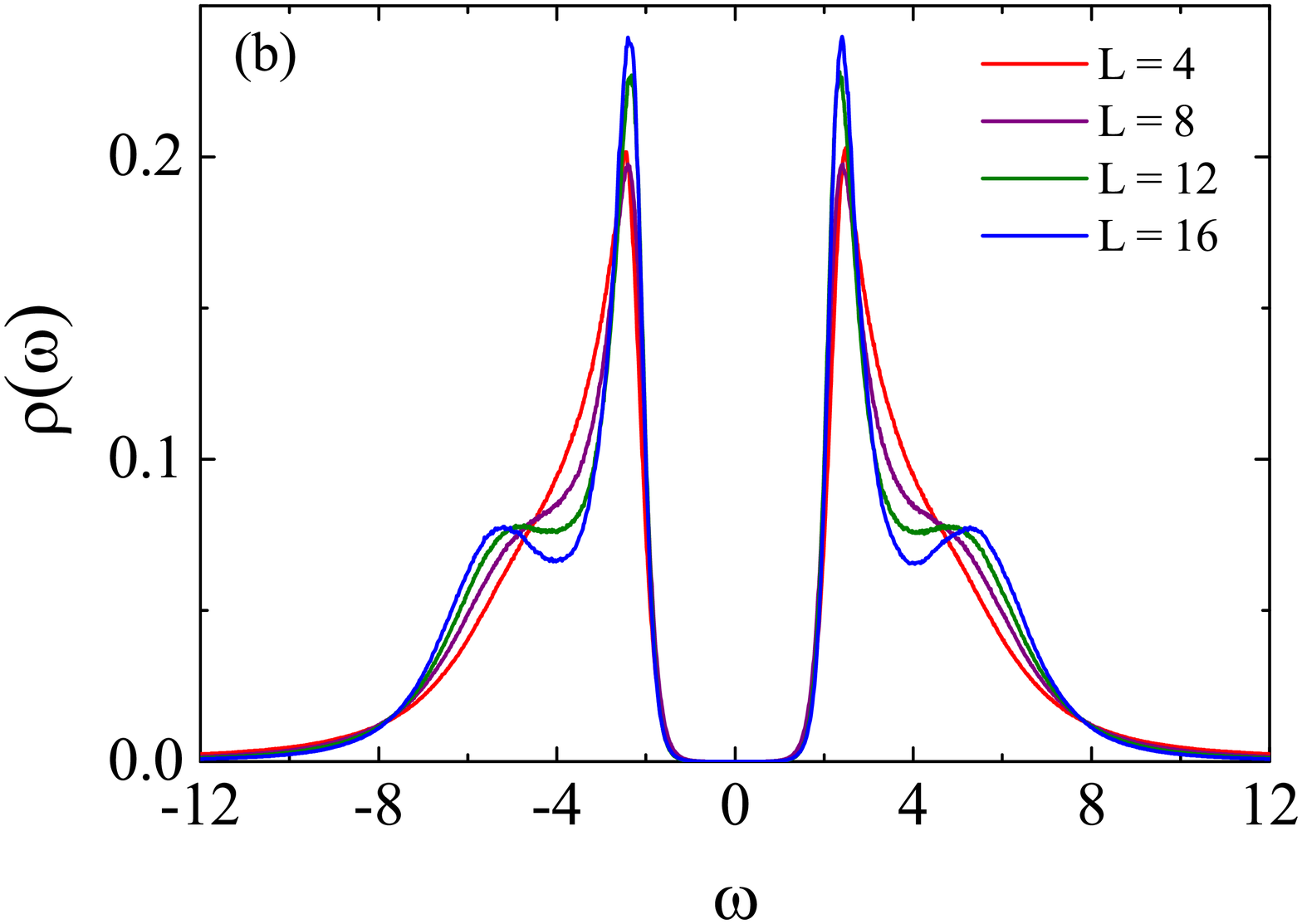}
\caption{QMC and analytic continuation at low temperatures:
(a) imaginary-time Green function, $G_{ii}(\protect\tau) = \frac{1}{N}
\sum_{\mathbf{k}\in\text{BZ}} G_\sigma (\mathbf{k},\protect\tau)$, at $U
 = 8$ and $\protect\beta = 8$ ($T = 0.125$) for $L = 4, \dots, 16$. The
logarithmic $y$-axis makes clear that $L = 8$, $12$, and $16$ give the same
slope in the imaginary-time decay, which ensures high-quality results on
analytic continuation. (b) Local density of states, $\protect\rho
(\protect \omega)$, obtained from SAC of the imaginary-time Green function
in panel (a). Results in the gap region have clearly converged for $L
 = 8$, $12$, and $16$ at this temperature.}
\label{fig:QMCdatab8}
\end{figure}

\begin{figure}[t]
\includegraphics[width=0.95\columnwidth]{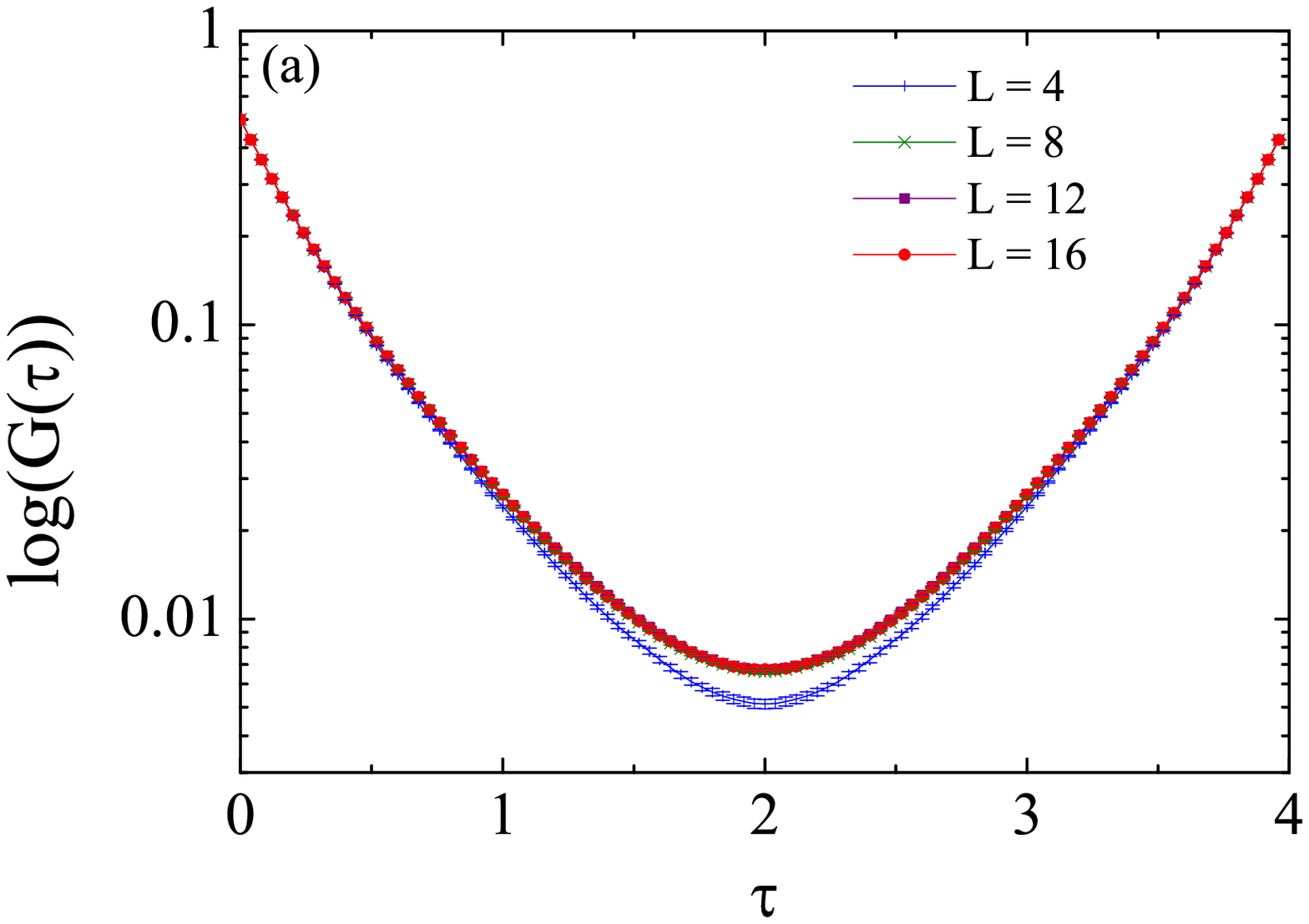}
\includegraphics[width=0.95\columnwidth]{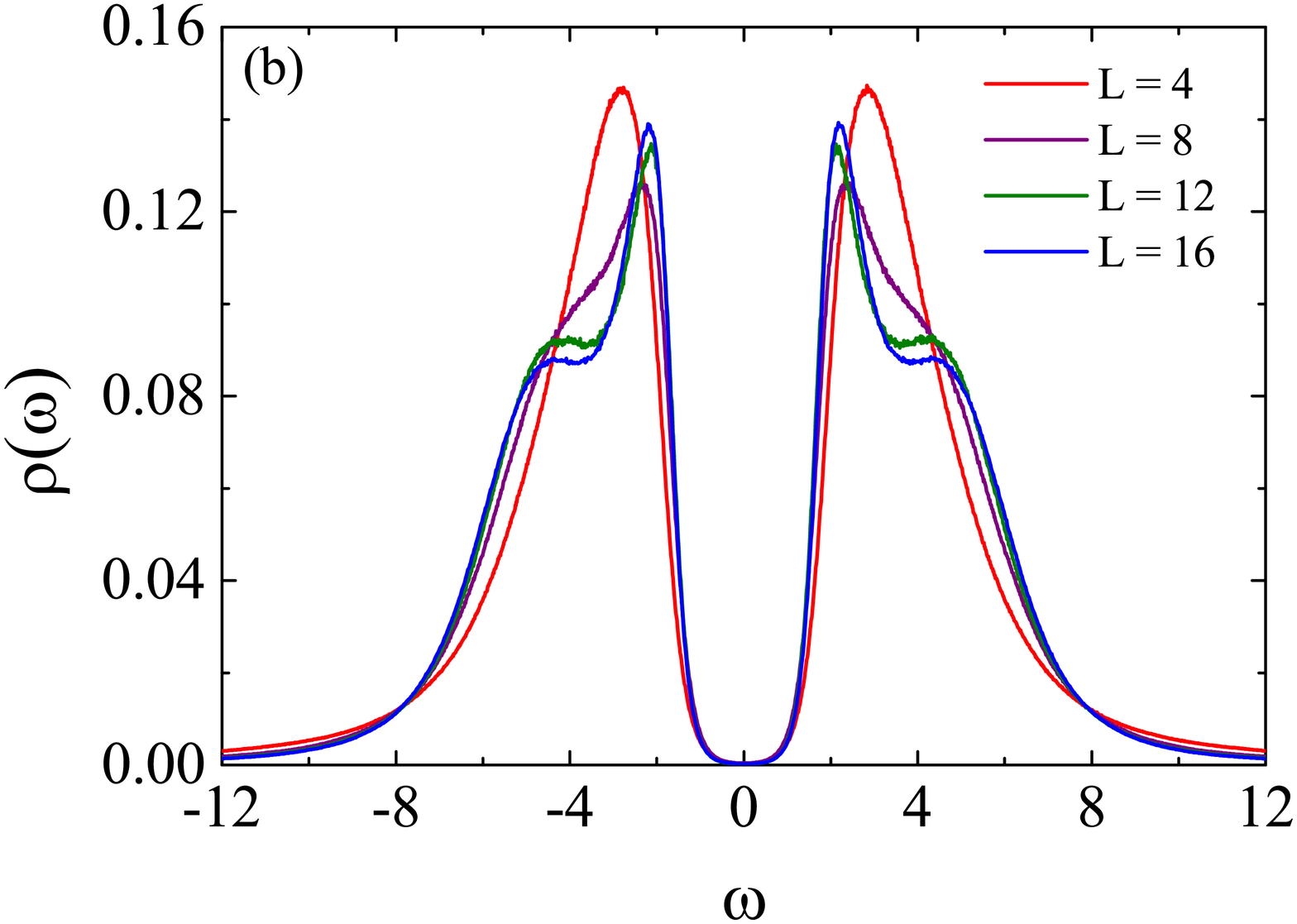}
\caption{QMC and analytic continuation at high temperatures:
(a) imaginary-time Green function, $G_{ii}(\protect\tau) =
\frac{1}{N} \sum_{\mathbf{k}\in\text{BZ}} G_\sigma (\mathbf{k},\protect\tau)$,
at $U = 8$ and $\protect\beta = 4$ ($T = 0.25$) for $L = 4, \dots, 16$.
Again the $L = 8$, $12$, and $16$ data provide converged values for the
imaginary-time decay, but an exponential form is no longer clear. (b)
Local density of states, $\protect\rho (\protect\omega)$, obtained from
SAC of the imaginary-time Green function in panel (a). The gap no longer
follows a single-exponential form, making it difficult to extract a reliable
value of $\Delta_{\text{Mott}}$ at this temperature.}
\label{fig:QMCdatab4}
\end{figure}

The QMC simulations give direct access to the imaginary-time fermion
Green function
\begin{equation}
G_{\sigma} (\mathbf{k},\tau) = - \frac{1}{N} \sum_{i,j}e^{i\mathbf{k} \cdot
(\mathbf{r}_{i} - \mathbf{r}_{j})} \langle c_{i\sigma } (\tau) c_{j\sigma}^{\dag} (0)
\rangle,
\end{equation}
where $i,j \in \lbrack 1,N]$ are site labels, $\tau \in \lbrack 0,\beta]$
is the imaginary time, and $\langle \dots \rangle $ is the Monte Carlo
expectation value. Concerning the spin index, $\sigma$, in the half-filled
Hubbard model $G(\mathbf{k},\tau) = G_{\uparrow}(\mathbf{k},\tau) = G_{\downarrow}
(\mathbf{k},\tau )$. While the slave-fermion treatment of Sec.~\ref{shmsf}
offers a specific calculation based on certain uncontrolled (but demonstrably
justified) approximations, the quantity $G(\mathbf{k},\tau)$ obtained from QMC
is exact on a finite-size system and has controlled errors.

To obtain real-frequency data, it is necessary to perform analytic
continuation of the imaginary-time data. For this purpose we have employed
stochastic analytic continuation (SAC) \cite{Beach-2004,Sandvik2016}, by which
the spectral function, $A(\mathbf{k},\omega)$, is obtained from the Green
function, $G(\mathbf{k},\tau)$, by a stochastic inverse Laplace transformation,
\begin{equation}
G(\mathbf{k},\tau) = \int d\omega \frac{e^{-\omega \tau }}{e^{-\beta \omega} + 1}
A(\mathbf{k},\omega).
\end{equation}
The most recent implementation of the SAC method reproduces the spectral
function using a large number of $\delta$-functions sampled at locations in a
frequency continuum and collected in a histogram \cite{Sandvik2016,YQQin2017,
HShao2017}. From the spectral function it is straightforward to obtain the
local density of states, $\rho (\omega) = \int_{\mathbf{k} \in \text{BZ}} d\mathbf{k}
A(\mathbf{k},\omega)$. Other static physical observables, such as the average
double site occupancy, $D = \frac{1}{N} \sum_{i} \langle n_{i\uparrow}
n_{i\downarrow} \rangle$, are also measured readily in QMC.

To access the single-particle gap, i.e.~the Mott gap ($\Delta_{\text{Mott}}$)
in what follows, one may attempt to read it directly from the gap in
$\rho(\omega)$. From the robust exponential decay of $G_{ii}(\tau)$ in
imaginary time at lower temperatures, shown for $\beta = 8$ in
Fig.~\ref{fig:QMCdatab8}(a), the analytic continuation is
straightforward and yields high-quality results for $\rho(\omega)$
[Fig.~\ref{fig:QMCdatab8}(b)]. We find in this temperature regime that the
density of states is well characterized by a single gap, $\Delta_{\text{Mott}}
 = 3.2(3)$. However, it becomes more difficult to extract an accurate value
for the Mott gap as the temperature increases. Figure \ref{fig:QMCdatab4}
shows $G_{ii}(\tau)$ and $\rho(\omega)$ at $U = 8$ but for $\beta = 4$ ($T
 = 0.25$). Although the imaginary-time decay of $G_{ii}(\tau)$ has converged
for $L = 8$, $12$, and $16$ [Fig.~\ref{fig:QMCdatab4}(a)], the finite-$T$
broadening that affects the Green function around $\tau = \beta/2$ makes
the fit to an exponential decay less accurate. From $\rho(\omega)$
[Fig.~\ref{fig:QMCdatab4}(b)], it remains clear at a qualitative level
that the spectrum has a gap, and that simulations for $L = 8$, $12$, and
$16$ converge to the same curve, but it is no longer clear how to ascribe
this behavior to a specific value of $\Delta_{\text{Mott}}$. We discuss
systematic ways of extracting lower and upper bounds on the Mott gap
from $\rho(\omega)$ in Sec.~\ref{sccs}C.

\begin{figure}[t]
\includegraphics[width=\columnwidth]{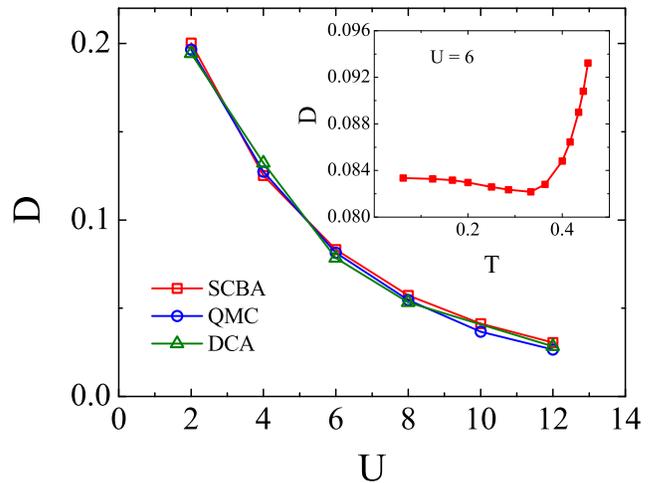}
\caption{Double occupancy, $D$, calculated as a function of $U$ at $T = 0.125$.
SCBA (red) and QMC (blue) results for 16$\times$16 systems are compared with
infinite-system results extrapolated from the dynamical cluster approximation
(DCA) \protect\cite{LeBlanc-2015} (green). Inset: $D(T)$ at $U = 6$ from SCBA.}
\label{fig:D_U}
\end{figure}

For a quantitative benchmarking of our SCBA results, we apply both methods on
systems of size 16$\times$16. Figure \ref{fig:D_U} shows the $U$-dependence of
the double occupancy at a temperature $T = 0.125$. $D$ reflects the extent of
charge fluctuations due to quantum and thermal effects. $D$ is suppressed as
$U$ increases, and we find excellent (percent-level) agreement of SCBA and
QMC. Also shown in Fig.~\ref{fig:D_U} are extrapolated DCA results
\cite{LeBlanc-2015}, which confirm not only the SCBA and QMC results
but also their convergence to the thermodynamic limit. The inset of
Fig.~\ref{fig:D_U} shows $D(T)$ computed at fixed $U = 6$. The weak
dip is a sensitive feature that has been the subject of extensive debate
\cite{Georges-1993,Werner-2005,Thereza-2010,Gorelik-2010,Imada-2016}. Our
slave-fermion approach provides a straightforward understanding of possible
nonmonotonic behavior in terms of the competition between weakening
spin-fluctuation-induced holon-doublon stabilization and strengthening
thermal fluctuations.

\begin{figure*}[tp]
\includegraphics[width=\textwidth]{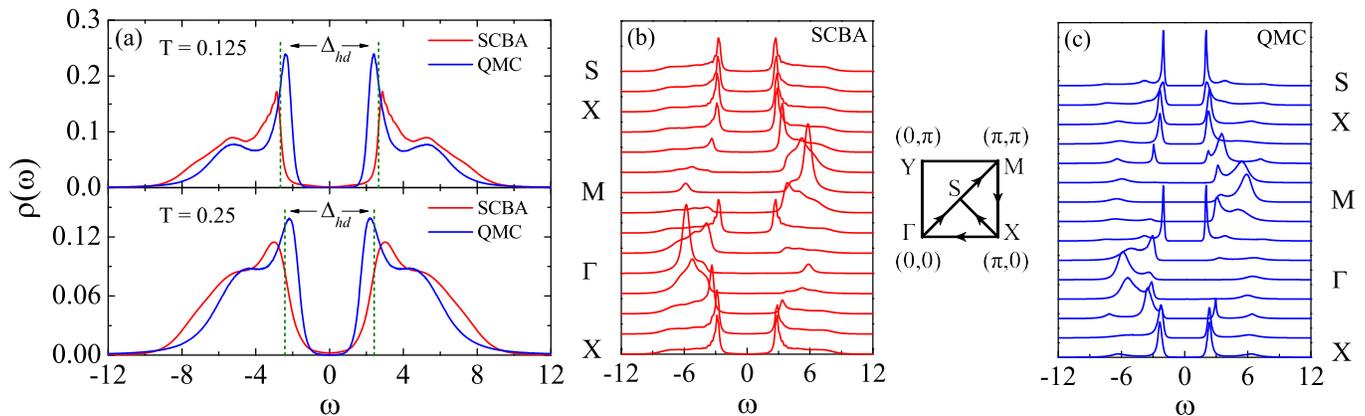}
\caption{(a) Electronic d.o.s., $\rho (\omega)$, computed for $U = 8$ by SCBA
(red) and QMC (blue) for $T = 0.125$ (upper panel) and $T = 0.25$ (lower).
Green dashed lines indicate the holon-doublon gap. (b) Spectral function,
$A(\mathbf{k},\omega)$, for $U = 8$ and $T = 0.125$, computed by SCBA and (c)
by QMC. (d) Path $\text{X} \rightarrow \Gamma \rightarrow \text{M} \rightarrow
\text{X} \rightarrow \text{S}$ of high-symmetry directions in the Brillouin
zone.}
\label{fig:DOS}
\end{figure*}

In the slave-fermion framework, the electron Green function, $G (\mathbf{k},
i\omega_{n})$, is the convolution of the charge (holon-doublon) and spin
propagators. Its calculation gives direct access to the electron spectral
function, $A(\mathbf{k},\omega) = - \frac{1}{\pi} \text{Im} \, G^{R}
(\mathbf{k},\omega + i \eta) $, and the density of states (d.o.s.), $\rho
(\omega) = \frac{1}{N} \sum_{\mathbf{k}} A (\mathbf{k},\omega)$. Figure
\ref{fig:DOS}(a) shows the SCBA and QMC d.o.s.~for $U = 8$ at $T = 0.125$
and 0.25. Three features are evident immediately. (i) Despite the absence
of AFM order, $\rho (\omega)$ shows a clear Mott (single-particle) gap.
$\Delta_{\text{Mott}}$, marking a region of strongly suppressed d.o.s.,
survives to $T > 0.25$, its decrease with $T$ signaling a ``melting'' of
the Mott insulator. (ii) The sharp peak at the Hubbard-band edge indicates
the emergence of a well-defined quasiparticle due to mutual charge and spin
renormalization. Following the discussion of a hole moving in an ordered AFM
\cite{Rink-1988,Kane-1989,Martinez-1991}, we name this feature a ``generalized
spin polaron'' and find that it loses coherence (as thermal fluctuations exceed
spin fluctuations) towards $T = 0.25$. (iii) At $T = 0.125$, $\rho (\omega)$
shows an obvious peak-dip-hump structure above the Mott gap, a much-debated
feature not captured in early QMC simulations \cite{Bulut-1994} but clearly
reproduced here by both SCBA and QMC.

Figures \ref{fig:DOS}(b) and \ref{fig:DOS}(c) show respectively
the SCBA and QMC spectral functions across the Brillouin zone for
$U = 8$. The results are again quantitatively similar in line shapes and
positions, albeit with differences in peak intensities and a small but
systematic discrepancy in gaps. The larger SCBA gaps may reflect an
overestimate of spin-fluctuation effects at intermediate $T$ values.

Extensive calculations of the type illustrated in Figs.~\ref{fig:D_U} and
\ref{fig:DOS} verify that the SCBA results are completely consistent with
QMC over the full range of intermediate $U$ and $T$. Thus it is safe to
conclude that the holon-doublon formulation and SCBA treatment do incorporate
correctly the interactions and mutual renormalization between the charge and
spin sectors. Hence the qualitative physics underlying the key features
of the Mott insulator, including the quasiparticle dynamics, Mott gap, and
pseudogap, can finally be uncovered.

\section{Convolution of Charge and Spin}
\label{sccs}

\begin{figure}[t]
\includegraphics[width=\columnwidth]{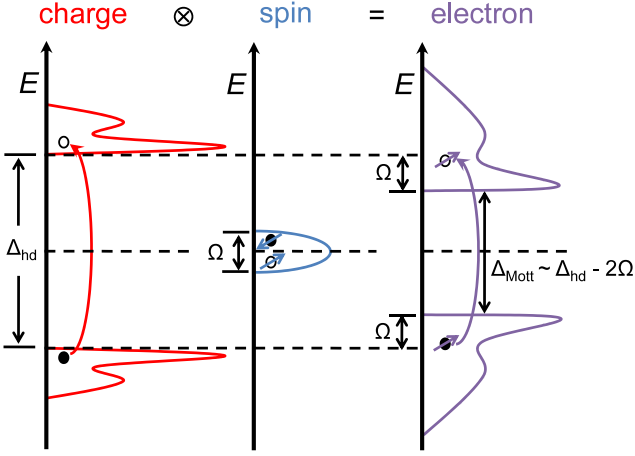}
\caption{Schematic representation of how the Mott-insulating state of
electrons (purple) is formed by convolution of the charge (red) and spin
(blue) degrees of freedom.}
\label{fs}
\end{figure}

More specifically, the slave-fermion framework makes it possible to separate
the contributions of the charge and spin sectors to the electronic spectral
function, which is a convolution of both. Figure \ref{fs} presents a
schematic illustration of the situation by showing an electronic excitation
in a Mott insulator. The lower and upper bands in the charge sector (red)
have the holon-doublon gap, $\Delta_{\text{hd}}$. This quantity defines the
high energy scale of the Mott insulator and its origin in holon-doublon
binding gives it a $T$-dependence analogous to the BCS superconducting gap.
In the spin sector (blue), low-energy particle-hole excitations exist over
a band of width $\Omega \approx
4J$, but only those on an energy scale $\Omega(T)$, which is governed by the
temperature, are activated. The electronic degrees of freedom (purple) are
reconstructed as the convolution of the two sectors and hence their excitations
are characterized by a gap $\Delta_{\text{Mott}}(T) \approx \Delta_{\text{hd}}(T)
 - 2 \Omega (T)$. In contrast to band insulators, where the gap is largely
$T$-independent \cite{Gebhard-1997}, the Mott gap is determined by
$T$-dependent correlation effects. As $T$ increases, $\Delta_{\text{Mott}}$ is
driven downwards both by the decrease in $\Delta_{\text{hd}}(T)$ and by the
increasing $\Omega(T)$.

\subsection{Charge Green function}
\label{scgf}

\begin{figure}[t]
\includegraphics[width=\columnwidth]{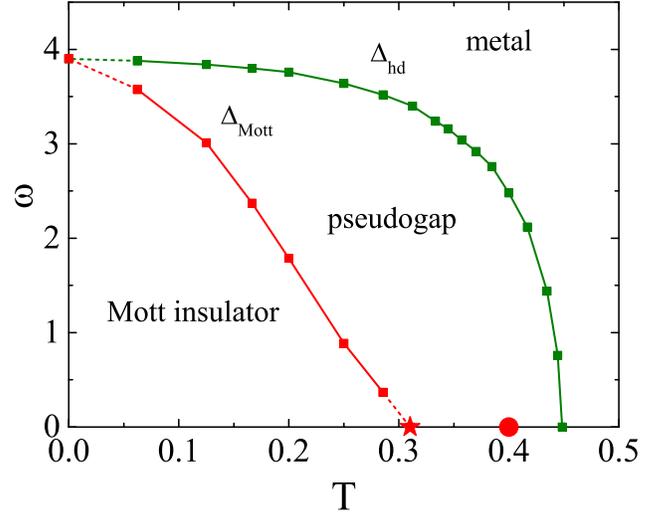}
\caption{Temperature-dependence of the charge ($\Delta_{\text{hd}}$, green)
and Mott ($\Delta_{\text{Mott}}$, red) gaps for the Hubbard model with $U = 6$,
estimated from the SCBA $\rho(\omega)$. Red squares indicate the lower bound
on $\Delta_{\text{Mott}}$. The red star and circle mark respectively the lower
and upper bounds on the temperature, $T_c^{\rm Mott}$, at which $\Delta_{\rm Mott}
(T)$ vanishes. Dashed red lines indicate extrapolations based on our
lower-bound values.}
\label{fg}
\end{figure}

We begin by considering the charge Green function of Eq.~(\ref{Charge}) in
order to extract $\Delta_{\text{hd}}(T)$. The dispersion relation, $E_{\mathbf{k}}$,
of the holon-doublon collective mode is obtained from the poles of this Green
function \cite{Mahan-1990} and the holon-doublon gap is twice its minimum
value, $\Delta_{\text{hd}} = 2 \min |_{\mathbf{k}} [|E_{\mathbf{k}}|]$. We exploit the
fact that the Pauli matrices, $\sigma_{i}$ ($i = 1$, 2, 3), and the identity
matrix, $I$, form a complete basis for all 2$\times$2 matrices to reexpress
Eq.~(\ref{Dis}) as
\begin{equation}
\tilde{\varepsilon}_{\mathbf{k}} = {\textstyle \frac{1}{2}} U \sigma _{3}
 - \zeta_{\mathbf{k}} \sigma_{2},
\end{equation}
where $\zeta_{\mathbf{k}}$ denotes $- 8itQ \eta_{\mathbf{k}}/J$. The self-energy
of the charge Green function (\ref{Charge}) is a 2$\times$2 matrix,
\begin{eqnarray}
\mathbf{\Sigma} (\mathbf{k},i\omega_{n}) & = & i \omega_{n} [1 - Z (\mathbf{k},
i\omega_{n})] I + \chi (\mathbf{k},i\omega_{n}) \sigma_{3} \notag \\ & & +
\phi_{1} (\mathbf{k},i\omega_{n}) \sigma_{1} + \phi_{2} (\mathbf{k},i\omega_{n})
\sigma_{2},
\label{Elia}
\end{eqnarray}
in which $Z (\mathbf{k},i\omega_{n})$ is the quasiparticle renormalization
factor, $\chi (\mathbf{k},i\omega_{n})$ contains the corrections to the
dispersion, and the off-diagonal terms, $\phi_{1} (\mathbf{k},i\omega_{n})$
and $\phi_{2} (\mathbf{k},i\omega_{n})$, contain the effects of the binding
interaction \cite{Eliashberg-1960,Scalapino-1966}. Substituting
Eq.~(\ref{Elia}) into Eq.~(\ref{Charge}) gives
\begin{equation}
\mathbf{F}^{-1} (\mathbf{k},i\omega_{n}) = \left( \!\! \begin{array}{cc}
\mathbf{F}_{11}^- (\mathbf{k},i\omega_{n}) &
 - \mathbf{F}_{12}^- (\mathbf{k},i\omega_{n}) \\
 - \mathbf{F}_{12}^+ (\mathbf{k},i\omega_{n}) &
\mathbf{F}_{11}^+ (\mathbf{k},i\omega_{n}) \end{array} \!\! \right) \!,
\label{EliaA}
\end{equation}
in which
\begin{eqnarray*}
\label{EliaB}
\mathbf{F}_{11}^{\pm} (\mathbf{k},i\omega_{n}) & = & Z (\mathbf{k},i\omega_{n})
i\omega_{n} \pm [U/2 + \chi (\mathbf{k},i\omega_{n})], \\ \mathbf{F}_{12}^{\pm}
(\mathbf{k},i\omega_{n}) & = & \phi_{1} (\mathbf{k},i\omega_{n}) \pm i [\phi_{2}
(\mathbf{k},i\omega _{n}) - \zeta_{\mathbf{k}}].
\end{eqnarray*}
By inversion of the matrix we obtain
\begin{equation}
\mathbf{F} (\mathbf{k},i\omega_{n}) = \frac{1}{|\text{DetF}|} \left( \!\!
\begin{array}{cc} \mathbf{F}_{11}^+ (\mathbf{k},i\omega_{n}) & \mathbf{F}_{12}^-
(\mathbf{k},i\omega_{n}) \\ \mathbf{F}_{12}^+ (\mathbf{k},i\omega_{n}) &
\mathbf{F}_{11}^- (\mathbf{k},i\omega_{n}) \end{array} \!\! \right),
\label{EliaC}
\end{equation}
where
\begin{eqnarray}
|\text{DetF}| & = & Z^{2} (\mathbf{k},i\omega_{n}) (i\omega_{n})^{2} - [U/2 +
\chi (\mathbf{k},i\omega_{n})]^{2} \nonumber \\ & & - \phi_{1}^{2} (\mathbf{k},
i\omega_{n}) - [\phi_{2} (\mathbf{k},i\omega_{n}) - \zeta_{\mathbf{k}}]^{2} \\
 & = & R (\mathbf{k},\omega) [\omega - E_{\mathbf{k}} + i \Gamma (\mathbf{k},
\omega)] \nonumber \\ & & \;\;\;\; \times [\omega + E_{\mathbf{k}} + i \Gamma
(\mathbf{k},\omega)]. \label{Final}
 \end{eqnarray}
We have calculated $\mathbf{F} (\mathbf{k},i\omega_{n})$ numerically, which
gives access to its component parts $Z (\mathbf{k},i\omega_{n})$, $\chi
(\mathbf{k},i\omega_{n})$, $\phi_{1} (\mathbf{k},i\omega_{n})$, and $\phi_{2}
(\mathbf{k},i\omega_{n})$. By reexpressing the denominator in the form given
in Eq.~(\ref{Final}), we derive the effective holon-doublon quasiparticle
dispersion, $E_{\mathbf{k}}$, and the corresponding scattering rate, $\Gamma
(\mathbf{k},\omega)$ \cite{Mahan-1990}. We
obtain the result for $\Delta_{\text{hd}}(T)$ shown as the green curve in
Fig.~\ref{fg}.

\subsection{Electron spectral function}
\label{sesf}

By contrast, $\Delta_{\text{Mott}}(T)$ is the single-particle gap and is smaller
than $\Delta_{\text{hd}}(T)$ due to the spin-renormalization of the charge sector
[Fig.~\ref{fs}]. This renormalization is contained in the slave-fermion
formulation within the convolution making up the electron Green function,
which if vertex corrections are neglected \cite{Raimondi-1993,Yamaji-2011}
may be expressed as
\begin{eqnarray}
G_{ij}^{\sigma}(\tau) & = & - \langle T_{\tau} c_{i\sigma}(\tau) c_{j\sigma}^{\dag}
(0) \rangle \nonumber \\ & = & - \langle T_{\tau} (s_{i\overline{\sigma}}^{\dagger}
(\tau) d_{i}(\tau) + \sigma e_{i}^{\dag}(\tau) s_{i\sigma }(\tau)) \nonumber \\
& & \;\;\;\;\;\; \times (d_{j}^{\dag}(0) s_{j\overline{\sigma }}(0) + \sigma
s_{j\sigma}^{\dag}(0) e_{j}(0)) \rangle \nonumber \\
& \simeq & - \langle T_{\tau} d_{i}(\tau) d_{j}^{\dag}(0) \rangle \langle
T_{\tau} s_{i\overline{\sigma}}^{\dag}(\tau) s_{j\overline{\sigma}}(0) \rangle
\\ & & - \langle T_{\tau} e_{i}^{\dag}(\tau) e_{j}(0) \rangle \langle T_{\tau}
s_{i\sigma}(\tau) s_{j\sigma}^{\dag}(0) \rangle \nonumber \\ & & - \sigma \langle
T_{\tau} d_{i}(\tau) e_{j}(0) \rangle \langle T_{\tau} s_{i\overline{\sigma}}^{\dag}
(\tau) s_{j\sigma}^{\dag}(0) \rangle \nonumber \\ & & - \sigma \langle T_{\tau}
e_{i}^{\dag}(\tau) d_{j}^{\dag}(0) \rangle \langle T_{\tau} s_{i\sigma}(\tau)
s_{j\overline{\sigma}}(0) \rangle. \nonumber
\end{eqnarray}
In momentum space it is given by
\begin{eqnarray}
G_{\sigma} & & (\mathbf{k},i\omega_{n}) \\ = & & \; \frac{1}{N} \sum_{\mathbf{q}}
\left( \int_{-\infty}^{\infty} d\varepsilon \, \frac{U_{\mathbf{q}}^{\dag} \mathbf{A}
(\mathbf{k+q},\varepsilon) U_{\mathbf{q}}}{i\omega_{n} + \Omega_{\mathbf{q}} -
\varepsilon} \, [f(\varepsilon) + n_{\mathbf{q}}] \right. \nonumber \\ & & \;\;
\left. + \int_{-\infty}^{\infty} d\varepsilon \, \frac{V_{\mathbf{q}}^{\dag} \mathbf{A}
(\mathbf{k+q},\varepsilon) V_{\mathbf{q}}}{i\omega_{n} - \Omega_{\mathbf{q}} -
\varepsilon } \, [1 - f(\varepsilon) + n_{\mathbf{q}}] \right) \!, \nonumber
\end{eqnarray}
with
\begin{equation}
U_{\mathbf{q}} = \left( \!\! \begin{array}{c} u_{\mathbf{q}} \\ v_{\mathbf{q}}^{\ast}
\end{array} \!\! \right) \;\; \text{and} \;\; V_{\mathbf{q}} = \left( \!\!
\begin{array}{c} v_{\mathbf{q}} \\ u_{\mathbf{q}}^{\ast} \end{array} \!\! \right) \!,
\end{equation}
whose components are given in Eq.~(\ref{euv}), and
\begin{eqnarray}
\mathbf{A} (\mathbf{k+q},\varepsilon) & = & - \frac{1}{\pi} \text{Im} \,
\mathbf{F}^{R} (\mathbf{k+q},\varepsilon + i\eta) \\ & = &
\left( \! \begin{array}{cc} A_{11} (\mathbf{k+q},\varepsilon) & A_{12}
(\mathbf{k+q},\varepsilon) \\ A_{21} (\mathbf{k+q},\varepsilon) & A_{22}
(\mathbf{k+q},\varepsilon) \end{array} \! \right) \!, \nonumber
\end{eqnarray}
which expresses the holon-doublon spectral function corresponding to the
retarded charge Green function; $f(\varepsilon)$ is the Fermi-Dirac
distribution function for holon-doublon quasiparticles and $n_{\mathbf{q}}$
the Bose-Einstein distribution for the spinons.

The corresponding electron spectral function is
\begin{eqnarray}
& & \tilde{A}_{\sigma} (\mathbf{k},\omega) = - \frac{1}{\pi} \textrm{Im} \,
G_{\sigma }^{R} (\mathbf{k},\omega) \\ & & = \; \frac{1}{N} \! \sum_{\mathbf{q}} \!\!
\int_{-\infty}^{\infty} \!\! d\varepsilon U_{\mathbf{q}}^{\dag} \mathbf{A}
(\mathbf{k \! + \! q},\varepsilon) U_{\mathbf{q}} [f(\varepsilon) \! + \!
n_{\mathbf{q}}] \delta(\omega \! + \! \Omega_{\mathbf{q}} \! - \! \varepsilon)
\notag \\ & & \!\! + \frac{1}{N} \! \sum_{\mathbf{q}} \!\! \int_{-\infty}^{\infty}
\!\!\! d\varepsilon V_{\mathbf{q}}^{\dag} \mathbf{A} (\mathbf{k \! + \! q},
\varepsilon) V_{\mathbf{q}} [1 \! - \! f(\varepsilon) \! + \! n_{\mathbf{q}}]
\delta (\omega \! - \! \Omega_{\mathbf{q}} \! - \! \varepsilon), \notag
\label{E_DOS}
\end{eqnarray}
whence the electronic density of states is
\begin{eqnarray}\label{cDOS}
\rho (\omega) & = & \frac{1}{N} \sum_{\mathbf{k},\sigma} \rho_{\sigma}
(\mathbf{k},\omega) \\ & = & \sum_{\mathbf{q}} a (\omega,\mathbf{q})
\int_{-\infty}^{\infty } d \varepsilon \rho_{hd}^{1} (\mathbf{q},\varepsilon)
\delta (\omega + \Omega_{\mathbf{q}} - \varepsilon) \notag \\ & & +
\sum_{\mathbf{q}} b (\omega,\mathbf{q}) \int_{-\infty}^{\infty} d \varepsilon
\rho_{hd}^{2} (\mathbf{q},\varepsilon) \delta (\omega - \Omega_{\mathbf{q}}
 - \varepsilon), \nonumber
\end{eqnarray}
in which
\begin{equation*}
\rho_{hd}^{1} (\mathbf{q},\varepsilon) = U_{\mathbf{q}}^{\dag} \mathbf{A}_{\mathbf{q}}
(\varepsilon) U_{\mathbf{q}}, \; \rho_{hd}^{2} (\mathbf{q},\varepsilon)
 = V_{\mathbf{q}}^{\dag} \mathbf{A}_{\mathbf{q}} (\varepsilon) V_{\mathbf{q}},
\end{equation*}
and
\begin{eqnarray*}
a(\omega,\mathbf{q}) & = & \frac{1}{N} [f(\omega + \Omega_{\mathbf{q}}) +
n_{\mathbf{q}}], \nonumber \\  b(\omega,\mathbf{q}) & = & \frac{1}{N}
[1 - f(\omega - \Omega_{\mathbf{q}}) + n_{\mathbf{q}}],
\end{eqnarray*}
where $\Omega_{\mathbf{q}}$ is the dispersion relation of the spin excitations.
The quantities $\rho_{hd}^{1} (\mathbf{q},\varepsilon)$ and $\rho_{hd}^{2}
(\mathbf{q},\varepsilon)$ contain the holon-doublon density of states
appearing in $\mathbf{A}_{\mathbf{q}} (\varepsilon)$, with nontrivial quantitative
modification by the spin part (contained in $U_{\mathbf{q}}$ and $V_{\mathbf{q}}$).
The temperature-dependence is contained within the occupation functions in
$a (\omega,\mathbf{q})$ and $b(\omega,\mathbf{q})$. The renormalization of
the holon-doublon gap to the Mott gap is contained within the integrals over
the two energy $\delta$-functions in Eq.~(\ref{cDOS}), $\delta (\omega \pm
\Omega_{\mathbf{q}} - \varepsilon)$, which mathematically effect the convolution
with the spin spectral function at the SCBA level and physically specify how
the lower and upper Hubbard bands are produced from holon-doublon bound states
dressed by the emission and absorption of low-energy spin fluctuations.

\begin{figure}[t]
\includegraphics[width=0.95\columnwidth]{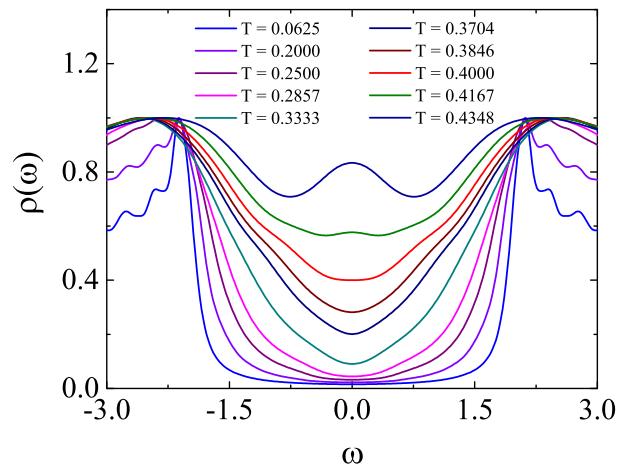}
\caption{$\rho (\omega)$ in the gap region at all temperatures, computed from
SCBA with $U = 6$ and $\eta = 0.1$; results are normalized to a peak height of
1 in order to highlight the emerging $\omega = 0$ peak.}
\label{fdos}
\end{figure}

\subsection{Extraction of the Mott gap}
\label{smge}

Unlike $\Delta_{\text{hd}}(T)$, the accurate extraction of $\Delta_{\text{Mott}}
(T)$ from the electron Green function is complicated by the lack of a
single-particle dispersion relation, and hence no analytical means of finding
the poles in the self-energy. However, as noted in Sec.~\ref{sec:QMC}, it is
even more difficult to read $\Delta_\text{Mott}$ from QMC for temperatures in
excess of approximately 0.15. Thus we revert to a detailed consideration of
the SCBA electronic d.o.s., $\rho (\omega,T)$ [Fig.~\ref{fdos}], in
order to estimate $\Delta_{\text{Mott}}(T)$ by a procedure of assuming an
effective gap and modelling its ``filling.'' We apply two types of analysis to
obtain ({\it 1}) a lower bound on $\Delta_{\rm Mott} (T)$, using a quantitative
fitting process which in essence neglects thermal fluctuations, and ({\it 2})
an upper bound on the temperature, $T_c^{\rm Mott}$, at which $\Delta_{\rm Mott}
(T) = 0$, based on a clear qualitative feature of $\rho (\omega,T)$.

\subsubsection{Lower bound on $\Delta_{\rm Mott}(T)$}

\begin{figure}[t]
\includegraphics[width=0.95\columnwidth]{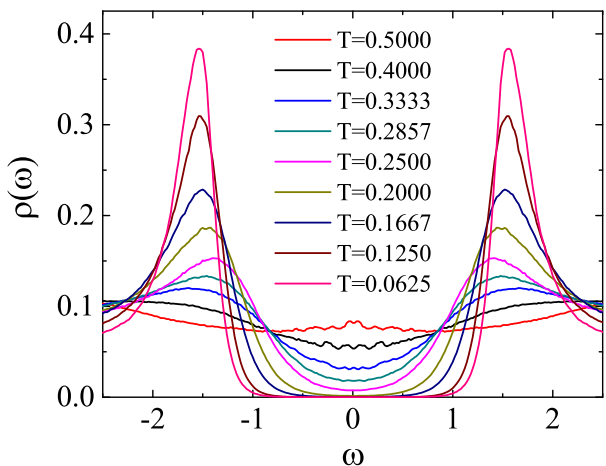}
\caption{$\rho (\omega)$ in the gap region for a number of temperature values,
computed by QMC with $U = 6$.}
\label{fig:QMC_U_6}
\end{figure}

Except at the highest temperatures, all of the d.o.s.~functions we
calculate by SCBA show the clear presence of a gap which, however, is
partially filled. This finite d.o.s.~at small $\omega$ is a consequence of
two effects, the (finite-size) Lorentzian broadening, $\eta$, and the
temperature, whose effects appear as exponential activation over an
effective $T$-dependent gap. A qualitative indication of the differing
nature of the two contributions can be obtained by comparing $\rho(\omega,T)$
from SCBA, shown in Fig.~\ref{fdos}, with the results from QMC, which we show
in Fig.~\ref{fig:QMC_U_6}: $\eta$ effects, which cause $\rho(\omega)$ to
become more ``V-shaped'' within the gap, are stronger in the SCBA data.
However, we are constrained by finite-size effects not to reduce $\eta$ in
our calculations. Because the Lorentzian contribution is much stronger, we
proceed by neglecting the thermal activation contribution, i.e.~the direct
effects of $T$, and thus obtain a lower bound for $\Delta_{\text{Mott}}(T)$.
Thus the problem of finding the Mott gap, meaning the gap in the
reconstructed (spin-charge-recombined) spectrum at any given $T$,
is reduced to a deconvolution removing $\eta$.

The retarded Green function can be represented by
\begin{equation}
G^{R} (\omega+i\eta) = \int_{-\infty}^{\infty} d\varepsilon \, \frac{{\tilde
\rho}(\varepsilon)}{\omega - \varepsilon + i\eta},
\label{Spec}
\end{equation}
where ${\tilde \rho}(\varepsilon)$, the intrinsic d.o.s., is expected to
vanish below $\Delta_{\text{Mott}}/2$. We need consider only the imaginary
part of $G^{R} (\omega + i\eta)$, which is the observed d.o.s.,
\begin{equation}
\rho (\omega) = \frac{1}{\pi} \int_{-\infty}^{\infty} d\varepsilon \, \frac{\eta}
{(\omega - \varepsilon)^2 + \eta^2} \, {\tilde \rho}(\varepsilon).
\label{Spec1}
\end{equation}
If $\eta$ is infinitesimal, at $T = 0$ and when the energy interval
is continuous one has ${\tilde \rho}(\omega) = \rho (\omega)$. In our
calculations, however, $\eta$ is finite and we have used an energy interval
$d\omega = 0.02$, on top of which we wish to demonstrate that the effect of
finite temperatures on the spectral function is equivalent to that of a
$T$-dependent effective Mott gap, $\Delta_{\rm Mott}(T)$.

\begin{figure}[t]
\includegraphics[width=0.96\columnwidth]{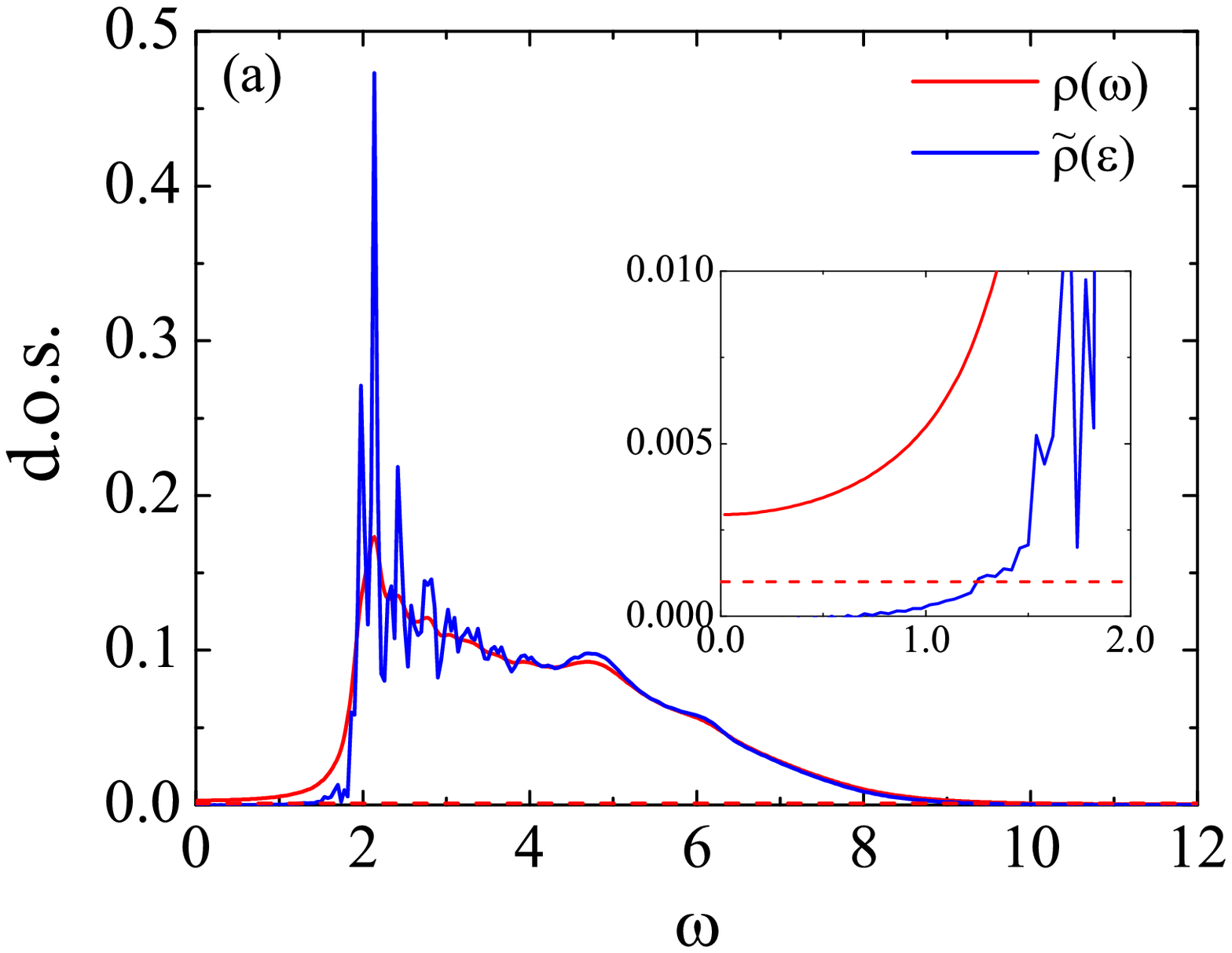}
\includegraphics[width=0.96\columnwidth]{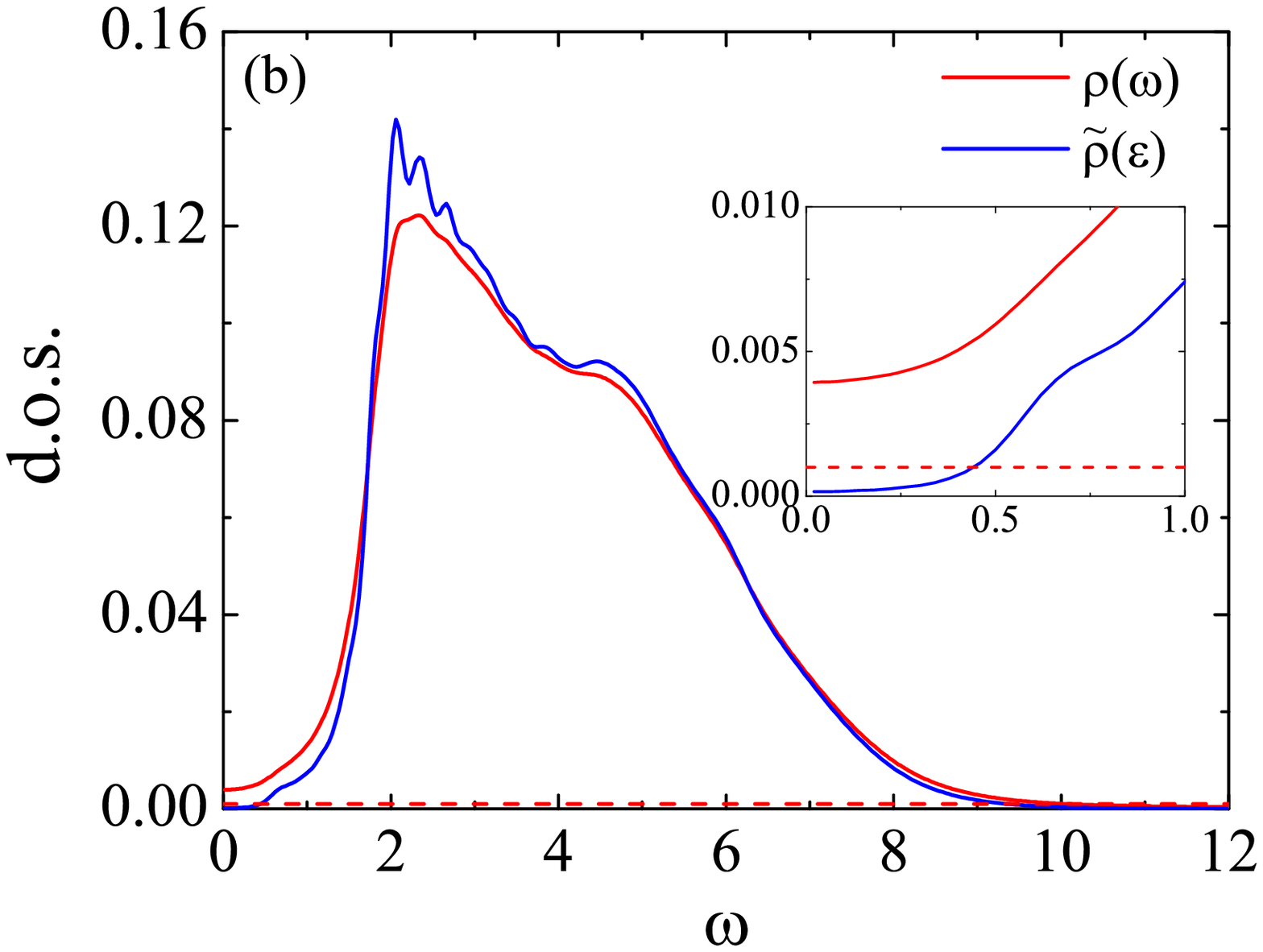}
\caption{Deconvolution of Lorentzian broadening in the SCBA d.o.s.
(a) Comparison of the functions $\rho(\omega)$, calculated by SCBA, and
$\tilde{\rho}(\varepsilon)$, obtained from it by linear regression, for $U
 = 6$ at temperature $T = 0.125$. (b) The same comparison at $T = 0.25$.
Clearly $\tilde{\rho}(\varepsilon)$ reveals additional intrinsic features
of the spectral function by removing the Lorentzian broadening, and hence
allows an estimate of $\Delta_{\text{Mott}}$. We comment that the SCBA data
for $\rho(\omega)$ contain $N_{\omega} = 600$ frequency points and the linear
regression is performed to obtain a dataset $\tilde{\rho}(\varepsilon)$
containing $N_{\varepsilon} = 300$ points. The dashed lines show the criterion
$\tilde{\rho}(\varepsilon) < 0.001$, on the basis of which we take the spectral
weight to vanish and thus define $\Delta_{\text{Mott}}$.}
\label{fig:SAC}
\end{figure}

As noted above, for a Mott insulator with no thermal fluctuations, one expects
that ${\tilde \rho} (\omega) = 0$ in the energy interval $[-\Delta_{\rm Mott}/2,
\Delta_{\rm Mott}/2]$, whence
\begin{equation}
\rho(\omega) = \frac{2}{\pi} \int_{\Delta_{\rm Mott}/2}^{\infty} d\varepsilon \,
\frac{\eta}{(\omega - \varepsilon)^2 + \eta^2} \, {\tilde \rho}
(\varepsilon).
\label{eq:M_S}
\end{equation}
The process of using $\rho(\omega)$, as calculated by SCBA at each value
of $T$, to extract the underlying function ${\tilde \rho} (\varepsilon)$
and the single constant $\Delta_{\rm Mott}(T)$ is analogous to an analytic
continuation. Although a full SAC treatment of the SCBA data is complicated by
a lack of statistical errors, a more straightforward procedure is sufficient
in the present case. Borrowing from the structure of the SAC method of
Sec.~\ref{sec:QMC}, we construct a minimization based on linear regression
to achieve the decomposition of Eq.~(\ref{eq:M_S}). We parameterize
\begin{equation}
\tilde{\rho}(\varepsilon) = \sum^{N_\epsilon}_{i=1} a_i \delta (\varepsilon -
\varepsilon_{i})
\label{eq:p}
\end{equation}
using $N_{\varepsilon}$ equally spaced $\delta$-functions, whose weights
$\{a_i\}$ are the free parameters. By inserting Eq.~(\ref{eq:p}) into
Eq.~(\ref{eq:M_S}), we obtain the function
\begin{equation}
\rho'(\omega) = \frac{1}{\pi} \sum^{N_{\epsilon}}_{i=1} d\varepsilon \left[
\frac{\eta}{(\omega - \varepsilon_i)^{2} + \eta^{2}} + \frac{\eta}
{(\omega + \varepsilon_i)^{2} + \eta^{2}} \right] \, a_{i},
\label{eq:linearregression}
\end{equation}
by which we approximate the SCBA $\rho(\omega)$. We define the goodness-of-fit
parameter
\begin{equation}
\chi^{2} = \sum_{i=1}^{N_{\omega}} \big( \rho (\omega_i) - \rho' (\omega_i)
\big)^2,
\end{equation}
whose minimization by a linear regression method determines the values
$\{a_i\}$. Because the number, $N_\varepsilon$, of data points in $\varepsilon$
in Eq.~(\ref{eq:p}) can only be equal to or smaller than the number, $N_\omega
 = 600$, of points in the SCBA $\rho(\omega)$, such a minimization can always
be achieved.

Two examples of the intrinsic d.o.s.~functions, $\tilde{\rho}(\varepsilon)$,
underlying our computed SCBA functions, $\rho(\omega)$, are shown in
Fig.~\ref{fig:SAC}, where we have chosen $U = 6$ and the temperatures
$T = 0.125$ [Fig.~\ref{fig:SAC}(a)] and $T = 0.25$ [Fig.~\ref{fig:SAC}(b)];
in both cases we used $N_{\varepsilon} = 300$. The $\tilde{\rho}(\varepsilon)$
functions show a clear suppression of the d.o.s.~at low frequencies, with
the reappearance of this weight occurring primarily around the peaks.
These intrinsic functions also show the clear presence of additional states
building systematically into the zero-temperature gap as $T$ is increased.

To extract the effective Mott gap from $\tilde{\rho}(\varepsilon)$ at each
temperature, we define $\Delta_{\mathrm{Mott}}(T)$ as the frequency at which
the weights $a_i$ in ${\tilde{\rho}} (\varepsilon)$ start to rise from zero.
More precisely, we use the criterion that $a_i$ should be less than 1\% of the
average d.o.s.~at the band center, $\rho(\omega) \approx 0.1$, i.e.~$a_i <
0.001$. As shown in the insets of Figs.~\ref{fig:SAC}(a) and \ref{fig:SAC}(b),
this criterion appears to offer a reliable means of distinguishing real
reconstructed finite-$T$ features from thermal and numerical noise.

By applying these considerations at $U = 6$, we obtain the data shown in
Fig.~\ref{fg}, with a well-defined lower bound from $T = 0.0625$ to
$T = 0.2857$. At our next higher temperature, $T = 0.333$, $a_i > 0.001$ even
at $\omega = 0$, and thus the lower bound has become zero; we estimate the
temperature at which this occurs to be $T \approx 0.31$, and represent this
by the dashed line in Fig.~\ref{fg}. We conclude that these results
can be taken to provide an accurate lower bound for $\Delta_{\rm Mott}(T)$, and
that the closing of the Mott gap by this estimate provides the lower bound,
$T_{c,l}^{\rm Mott} \approx 0.31$, for the associated temperature.

\subsubsection{Upper bound on $T_c^{\rm Mott}$}

A qualitatively different approach to the Mott transition is provided by the
fact that, as Figs.~\ref{fdos} (SCBA) and \ref{fig:QMC_U_6} (QMC) make
clear, $\rho(\omega)$ changes from a low-$T$ form with an absolute minimum at
$\omega = 0$ to a high-$T$ form with a peak at $\omega = 0$. This peak grows
in size and spectral weight as a function of temperature beyond a given $T$
value. The emergence of this quasiparticle peak in the single-particle response
indicates unambiguously that the lower and upper Hubbard bands have overlapped,
and thus that the Mott gap has closed. In practice, $\Delta_{\text{Mott}}$ may
have vanished before the peak can emerge as a feature stronger than the
d.o.s.~at neighboring nonzero frequencies, and hence this temperature,
$T_{c,u}^{\rm Mott}$, can be taken as an upper bound for the closing of the
Mott gap. At $U = 6$ we find, as shown in Fig.~\ref{fg}, that
$T_{c,u}^{\rm Mott} \approx 0.4$. Thus both $T_{c,l}^{\rm Mott}$ and $T_{c,u}^{\rm Mott}$
lie well below the closing temperature of the holon-doublon gap, $T^{\text{hd}}_c
\simeq 0.45$. While the QMC results differ from SCBA in quantitative details,
the qualitative picture of the two gaps remains robust.

\section{Interpretation}
\label{si}

The $\omega = 0$ axis of Fig.~\ref{fg} can be interpreted as
a finite-temperature phase diagram for the Hubbard model (\ref{ehm}). The
low-$T$ regime is fully gapped, not only for bound holons and doublons but
also for electrons, and this is the Mott insulator. As $T$ is increased,
its melting is revealed as a two-step process. At $T_c^{\text{Mott}}$, the
optimal electronic states created by spin-charge reconstruction, which lie
in the low-energy tails of the Hubbard bands, have touched, creating the
$\omega = 0$ peak in $\rho (\omega)$ [Fig.~\ref{fdos}]. However,
$\Delta_{\text{hd}}(T)$ remains finite and most electronic states remain gapped;
the consequent suppression of the d.o.s.~around the Fermi level makes this a
pseudogap regime. As $T$ approaches $T_c^{\text{hd}}$, the pseudogap fills in with
additional low-lying electronic states, and only above $T_c^{\text{hd}}$ does
the closing of the charge gap make the system fully metallic.

This pseudogap behavior \cite{White-1993} is observed consistently in many
numerical studies of the Hubbard model \cite{LeBlanc-2015}, including our own
(Secs.~\ref{sec:QMC} and \ref{sccs}C), and is not restricted to finite doping.
The slave-fermion framework captures this phenomenon, showing how the
single-particle gap decreases faster than the charge gap before the
vanishing of both quantities establishes the two characteristic temperatures,
$T_c^{\rm Mott}$ and $T_c^{\rm hd}$ [Fig.~\ref{fg}]. Even at low $T$, the
difference between $\Delta_{\text{hd}}$ and $\Delta_{\text{Mott}}$ grows linearly
with $T$, as anticipated in Sec.~\ref{sccs}. This difference in the
$T$-dependences of the two gaps constitutes one crucial feature of the
distinctive low-energy physics intrinsic to the Mott insulator.

A further piece of essential physics concerns the energy scales of the
spin fluctuations and gap renormalization. Energies in the spin sector are
controlled by $J$, and the relevant temperatures for a finite (two-spin)
magnetic correlation parameter are a fraction of this. However, charge-sector
energies are of order $U$, and the renormalization, $\Omega(T)$, of
$\Delta_{\text{hd}} (T)$ is a fraction of this. From Fig.~\ref{fg},
$\Omega(T) \approx 5 T \approx U T$. This remarkable ``leverage
effect,'' by which the low-energy spin processes bring about high energy
shifts in the charge processes, lies at the heart of the energy-scale mixing
in the Mott insulator.

Equations (\ref{cDOS}) quantify how effectively the spin part modifies the
filling of the charge gap during the reconstruction of the electronic degrees
of freedom, and as such the leverage effect is not a directly modified energy
scale, but rather a strong amplification of the gap-filling. When this process
is modeled as a thermal filling governed by an effective (Mott) gap for the
electrons (Sec.~\ref{sccs}C), one finds the factor of order $U$ that enters
in the difference from the charge gap. Thus the mixing of energy scales
intrinsic to the Mott insulator is revealed from a set of equations affording
physical insight, rather than ``only'' as the output from a complex numerical
simulation.

These same equations also provide a quantitative description of the spectral
weight in the pseudogap regime. Although the presence of particularly low-lying
reconstructed electronic states has already closed the Mott gap, these states
constitute only a small fraction of the total available electronic states,
the majority of which still reside above the charge gap. As a consequence,
the electronic d.o.s.~in the low-energy regime remains suppressed, which is
the definition of the pseudogap, and this suppression persists until the
temperature is high enough to close the charge gap. We comment that, as in
Ref.~\cite{Han-2016}, the electronic Green function may also be used to
obtain the Luttinger surface for the insulating regime and the Fermi surface
of the pseudogap and metallic states; however, because the model we consider
is half-filled and has no extended hopping, this surface is precisely the
Umklapp surface (the locus of solutions of $\cos k_x + \cos k_y = 0$) for
all temperatures.

\section{Concluding remarks}
\label{scr}

To place our results in context, all slave-particle decompositions involve
an uncontrolled assumption, which is justified {\it post facto}. Our QMC
simulations reveal that the holon-doublon approach does an excellent job of
representing the relevant degrees of freedom and of capturing all the important
aspects of their interactions. Further, any mean-field treatment enforces the
local constraint only on average, making its results critically dependent on
how well the essential physics is captured at lowest order. Again the
holon-doublon framework passes this test with distinction, for all values of
$U > 2 t$ [Fig.~\ref{fig:D_U}] and temperatures $T \lesssim 0.5 J$. Unlike
some approaches, our study is general in that the finite-$T$ response contains
no potential pathologies induced by the perfectly nested noninteracting band.

Experimentally, despite intensive interest in cuprate materials and Mott
physics, detailed studies of undoped Mott insulators are complicated in
that neither angle-resolved photoemission spectroscopy (ARPES) nor scanning
tunnelling spectroscopy (STS) can obtain a signal from a well-gapped insulator
at low $T$. ARPES on insulating cuprates \cite{Damascelli-2003} has mapped the
spectral function [Fig.~\ref{fig:DOS}(b)] to observe the Mott gap and strongly
renormalized bands, but lack the resolution and temperature-sensitivity to
address the filling and closing of $\Delta_{\text{Mott}}$. STS measures the local
d.o.s.~[Fig.~\ref{fig:DOS}(a)] and recent studies \cite{rw1,rw2} have observed
the Mott gap and its persistence to finite temperatures, albeit in systems
that are already lightly hole-doped (which is the next challenge for the
slave-fermion description). Very recently, AFM order has been observed in
ultracold $^6$Li atoms on an optical lattice, which also realize an undoped
Hubbard model at finite temperatures \cite{rca}. Given the finite nature (of
order 100 atoms) of these systems, our SCBA and QMC techniques are both
perfectly suited for calculations and quantitative comparison with this
type of experiment.

In summary, we have shown by analytical SCBA calculations and unbiased
QMC simulations that the slave-fermion description of the Hubbard model
contains all the essential physics of the Mott insulator. Thus we obtain
complete insight into the underlying physical processes, which emerge from
high-energy holon-doublon binding mediated by low-energy, short-ranged spin
fluctuations. The spin-renormalization of the charge sector, which forms the
lower and upper Hubbard (electron) bands, introduces a strong energetic
leverage effect. The Mott gap is smaller than the charge gap and its closure
involves only a fraction of the reconstructed states, giving a natural
explanation of the pseudogap. This unified understanding of the dynamics and
melting of the undoped Mott insulator forms a sound basis for investigating
both the doped case and specific cuprate band structures.

\begin{acknowledgments}

We thank A.~W.~Sandvik and R.~Yu for helpful
discussions. This work was supported by the National Natural Science
Foundation of China (Grant Nos.~10934008, 10874215, 11174365, and 11574359),
by the National Basic Research Program of China (Grant Nos.~2012CB921704,
2011CB309703, and 2016YFA0300502) and by the Chinese Academy of Sciences
under Grant No.~XDPB0803.

\end{acknowledgments}


\begin{thebibliography}{99}

\bibitem{Mott-1937} N. F. Mott and R. Peierls, Proc. Phys. Soc. London
\textbf{49}, 72 (1937).

\bibitem{Mott-1949} N. F. Mott, Proc. Phys. Soc., London, Sect. A \textbf{62},
416 (1949).

\bibitem{Mott-1956} N. F. Mott, Can. J. Phys. \textbf{34}, 1356 (1956).

\bibitem{Imada-1998} M. Imada, A. Fujimori, and Y. Tokura, Rev. Mod. Phys.
\textbf{70}, 1039 (1998).

\bibitem{Bednorz-1986} J. G. Bednorz and K. A. M\"{u}ller, Z. Phys. B:
Condens. Matter \textbf{64}, 189 (1986).

\bibitem{Lee-2006} P. A. Lee, N. Nagaosa, and X.-G. Wen, Rev. Mod. Phys.
\textbf{78}, 17 (2006).

\bibitem{Damascelli-2003} A. Damascelli, Z. Hussain, and Z.-X. Shen, Rev.
Mod. Phys. \textbf{75}, 473 (2003).

\bibitem{Norman-2005} M. R. Norman, D. Pines, and C. Kallin, Adv. Phys.
\textbf{54}, 715 (2005).

\bibitem{Sawatzky-2008} S. Hufner, M. A. Hossain, A. Damascelli, and G. A.
Sawatzky, Rep. Prog. Phys. \textbf{71}, 062501 (2008).

\bibitem{Mott-1990} N. F. Mott, \textit{Metal-Insulator Transitions} (Taylor
and Francis, London, 1990).

\bibitem{Hubbard-1963} J. Hubbard, Proc. Roy. Soc. London, Ser. A \textbf{276},
238 (1963).

\bibitem{Brinkman-1970} W. F. Brinkman and T. M. Rice, Phys. Rev. B \textbf{2},
4302 (1970).

\bibitem{Huscroft-2001} C. Huscroft, M. Jarrell, Th. Maier, S. Moukouri, and
A. N. Tahvildarzadeh, Phys. Rev. Lett. \textbf{86}, 139 (2001).

\bibitem{Moukouri-2001} S. Moukouri and M. Jarrell, Phys. Rev. Lett. \textbf{%
87}, 167010 (2001).

\bibitem{Kyung-2006} B. Kyung, S. S. Kancharla, D. S\'{e}n\'{e}chal, A.-M.
S. Tremblay, M. Civelli, and G. Kotliar, Phys. Rev. B \textbf{73}, 165114
(2006).

\bibitem{Park-2008} H. Park, K. Haule, and G. Kotliar, Phys. Rev. Lett.
\textbf{101}, 186403 (2008).

\bibitem{Sordi-2010} G. Sordi, K. Haule, and A.-M. S. Tremblay, Phys. Rev.
Lett. \textbf{104}, 226402 (2010).

\bibitem{Gunnarsson-2015} O. Gunnarsson, T. Sch\"{a}fer, J. P. F. LeBlanc,
E. Gull, J. Merino, G. Sangiovanni, G. Rohringer, and A. Toschi, Phys. Rev.
Lett. \textbf{114}, 236402 (2015).

\bibitem{Castellani-1979} C. Castellani, C. Di Castro, D. Feinberg, and J.
Ranninger, Phys. Rev. Lett. \textbf{43}, 1957 (1979).

\bibitem{Kaplan-1982} T. A. Kaplan, P. Horsch, and P. Fulde, Phys. Rev.
Lett. \textbf{49}, 889 (1982).

\bibitem{Capello-2005} M. Capello, F. Becca, M. Fabrizio, S. Sorella, and
E. Tosatti, Phys. Rev. Lett. \textbf{94}, 026406 (2005).

\bibitem{Yokoyama-2006} H. Yokoyama, M. Ogata, and Y. Tanaka, J. Phys. Soc.
Jpn. \textbf{75}, 114706 (2006).

\bibitem{Phillips-2010} P. Phillips, Rev. Mod. Phys. \textbf{82}, 1719
(2010).

\bibitem{Sen-2014} S. Zhou, Y. Wang, and Z. Wang, Phys. Rev. B
\textbf{89}, 195119 (2014).

\bibitem{Peter-2015} P. Prelov\v{s}ek, J. Kokalj, Z. Lenar\v{c}i\v{c}, and
R. H. McKenzie, Phys. Rev. B \textbf{92}, 235155 (2015).

\bibitem{Han-2016} X.-J. Han, Y. Liu, Z.-Y. Liu, X. Li, J. Chen, H.-J. Liao,
Z.-Y. Xie, B. Normand, and T. Xiang, New J. Phys. \textbf{18}, 103004 (2016).

\bibitem{Kotliar-1986} G. Kotliar and A. E. Ruckenstein, Phys. Rev. Lett.
\textbf{57}, 1362 (1986).

\bibitem{LeBlanc-2015} J. P. F. LeBlanc, A. E. Antipov, F. Becca, I. W.
Bulik, G. K.-L. Chan, C.-M. Chung, Y. Deng, M. Ferrero, T. M. Henderson,
C. A. Jim{\'e}nez-Hoyos, E. Kozik, X.-W. Liu, A. J. Millis, N. V. Prokof’ev,
M. Qin, G. E. Scuseria, H. Shi, B. V. Svistunov, L. F. Tocchio, I. S. Tupitsyn,
S. R. White, S. Zhang, B.-X. Zheng, Z. Zhu, and E. Gull, Phys. Rev. X.
\textbf{5}, 041041 (2015), and references therein.

\bibitem{Auerbach-1994} A. Auerbach, \textit{Interacting Electrons and
Quantum Magnetism} (Springer-Verlag, New York, 1994).

\bibitem{Arovas-1988} D. P. Arovas and A. Auerbach, Phys. Rev. B \textbf{38},
316 (1988).

\bibitem{Yoshioka-1989} D. Yoshioka, J. Phys. Soc. Jpn. \textbf{58}, 1516
(1989).

\bibitem{rwscsgf} W. Wu, M. S. Scheurer, S. Chatterjee, S. Sachdev, A.
Georges, and M. Ferrero, Phys. Rev. X {\bf 8}, 021048 (2018).

\bibitem{rscwfgs}
M. S. Scheurer, S. Chatterjee, W. Wu, M. Ferrero, A. Georges, and S. Sachdev,
Proc. Natl. Acad. Sci. USA {\bf 115}, E3665 (2018).

\bibitem{Manousakis-1991} E. Manousakis,
Rev. Mod. Phys. \textbf{63}, 1 (1991).

\bibitem{Chakravarty-1989} S. Chakravarty, B. I. Halperin, and D. R. Nelson,
Phys. Rev. B \textbf{39}, 2344 (1989).

\bibitem{Schulz-1990} H. J. Schulz,
Phys. Rev. Lett. \textbf{65}, 2462 (1990).

\bibitem{Dupuis-2004} K. Borejsza and N. Dupuis,
Phys. Rev. B \textbf{69}, 085119 (2004).

\bibitem{BSS-1981} R. Blankenbecler, D. J. Scalapino, and R. L. Sugar, Phys.
Rev. D \textbf{24}, 2278 (1981).

\bibitem{Hirsch-1983} J. E. Hirsch, Phys. Rev. B \textbf{28}, 4059(R) (1983).

\bibitem{Hirsch-1985} J. E. Hirsch, Phys. Rev. B \textbf{31}, 4403 (1985).

\bibitem{Scalettar-1991} R. T. Scalettar, R. M. Noack, and R. R. P. Singh,
Phys. Rev. B \textbf{44}, 10502 (1991).

\bibitem{quest} The finite-temperature QMC simulation code is based on the
Quantum Electron Simulation Toolbox (QUEST), which is a FORTRAN 90/95 package
containing modern algorithms, such as delayed updating, and integrating the
latest BLAS/LAPACK numerical kernels. QUEST has integrated several legacy
codes by modularizing their computational components for ease of maintenance
and program interfacing. The current version can be accessed at
https://code.google.com/archive/p/quest-qmc/.

\bibitem{Beach-2004} K. S. D. Beach, unpublished (arXiv:cond-mat/0403055).

\bibitem{Sandvik2016} A. W. Sandvik, Phys. Rev. E \textbf{94}, 063308 (2016).

\bibitem{YQQin2017} Y. Q. Qin, B. Normand, A. W. Sandvik, and Z. Y. Meng,
Phys. Rev. Lett. \textbf{118}, 147207 (2017).

\bibitem{HShao2017} H. Shao, Y. Q. Qin, S. Capponi, S. Chesi, Z. Y. Meng, and
A. W. Sandvik, Phys. Rev. X \textbf{7}, 041072 (2017).

\bibitem{Georges-1993} A. Georges and W. Krauth, Phys. Rev. B \textbf{48},
7167 (1993).

\bibitem{Werner-2005} F. Werner, O. Parcollet, A. Georges, and S. R. Hassan,
Phys. Rev. Lett. \textbf{95}, 056401 (2005).

\bibitem{Thereza-2010} T. Paiva, R. Scalettar, M. Randeria, and N. Trivedi,
Phys. Rev. Lett. \textbf{104}, 066406 (2010).

\bibitem{Gorelik-2010} E. V. Gorelik, I. Titvinidze, W. Hofstetter, M.
Snoek, and N. Bl\"{u}mer, Phys. Rev. Lett. \textbf{105}, 065301 (2010).

\bibitem{Imada-2016} K. Takai, K. Ido, T. Misawa, Y. Yamaji, and M. Imada,
J. Phys. Soc. Jpn. \textbf{85}, 034601 (2016).

\bibitem{Rink-1988} S. Schmitt-Rink, C. M. Varma, and A. E. Ruckenstein,
Phys. Rev. Lett. \textbf{60}, 2793 (1988).

\bibitem{Kane-1989} C. L. Kane, P. A. Lee, and N. Read, Phys. Rev. B
\textbf{39}, 6880 (1989).

\bibitem{Martinez-1991} G. Martinez and P. Horsch, Phys. Rev. B \textbf{44},
317 (1991).

\bibitem{Bulut-1994} N. Bulut, D. J. Scalapino, and S. R. White, Phys. Rev.
Lett. \textbf{73}, 748 (1994).

\bibitem{Gebhard-1997} F. Gebhard, \textit{The Mott metal-insulator
transition: Models and methods} (Springer, Berlin, 1997).

\bibitem{Mahan-1990} G. D. Mahan, \textit{Many-Particle Physics} (Plenum
Press, New York, 1990).

\bibitem{Eliashberg-1960} G. M. Eliashberg, Sov. Phys. JETP \textbf{11}, 696
(1960).

\bibitem{Scalapino-1966} D. J. Scalapino, J. R. Schrieffer, and J. W. Wilkins,
Phys. Rev \textbf{148}, 263 (1966).

\bibitem{Raimondi-1993} R. Raimondi and C. Castellani,
Phys. Rev. B \textbf{48}, 11453 (1993).

\bibitem{Yamaji-2011} Y. Yamaji and M. Imada,
Phys. Rev. B \textbf{83}, 214522 (2011).

\bibitem{White-1993} M. Veki\'{c} and S. R. White, Phys. Rev. B \textbf{47},
1160(R) (1993).

\bibitem{rw1} W. Ruan, C. Hu, J. Zhao, P. Cai, Y. Peng, C. Ye, R. Yu, X. Li,
Z. Hao, C. Jin, X. Zhou, Z.-Y. Weng, and Y. Wang, Science Bull. {\bf 61},
1826 (2016).

\bibitem{rw2} P. Cai, W. Ruan, Y. Peng, C. Ye, X. Li, Z. Hao, X. Zhou, D.-H.
Lee, and Y. Wang, Nature Phys. {\bf 12}, 1047 (2016).

\bibitem{rca} A. Mazurenko, C. S. Chiu, G. Ji, M. F. Parsons, M. Kan\'asz-Nagy,
R. Schmidt, F. Grusdt, E. Demler, D. Greif, and M. Greiner, Nature {\bf 545},
462 (2017).

\end{thebibliography}
\end{document}